# Fine-tuned LLM-based Code Migration Framework

Oleg Grynets
EPAM Systems
McLean, Virginia, USA
oleg_grynets@epam.com

Vasyl Lyashkevych
EPAM Systems
Lviv, Ukraine
vasyl_lyashkevych@epam.com

Dmytro Baran
EPAM Systems
Lviv, Ukraine
dmytro_baran@epam.com

Maksym Orliansky
EPAM Systems
Lviv, Ukraine
maksym_orliansky@epam.com

Taras Zelenyy
EPAM Systems
Lviv, Ukraine
taras_zelenyy@epam.com

Markiian Leshchyshyn
EPAM Systems
Lviv, Ukraine
markiian_leshchyshyn@epam.com

***Abstract*—This study presents the outcomes of research and experimental validation in the domain of automated codebase migration, with a focus on addressing challenges in transitioning SQL-based systems, specifically from Oracle to PostgreSQL. The proposed method for migration essentially appears as a framework that leverages the best aspects of traditional software engineering techniques with cutting-edge Generative AI (GAI) technologies. It provides an iterative, scalable, precise and efficient solution for modern database transformations. The iterative nature of the method is ensured by the approach to dataset preparation, a flexible set of metrics, and a deep analysis of conversion errors and their distribution among the features of the SQL codebase. This allows us to predict quite accurately what percentage of the code and with what quality it will be transformed. The central piece of the approach is the integration of a fine-tuned Large Language Model (LLM) to address critical issues in SQL code conversion, such as syntax mapping, resolving discrepancies between Oracle PL/SQL and PostgreSQL, and optimising database elements such as stored procedures, triggers, views, and overall database logic. Thus, the method involves a trade-off between fine-tuning and prompt engineering. Special attention is given to a fine-tuning approach, which enhances the adaptability and compatibility with migration requirements across the entire database. According to the achieved results, fine-tuning plays a very important role. The study employs targeted evaluation methodologies along with computational metrics to measure the success of iterative conversion cycles. Core innovations include automated SQL feature detection, semi-supervised error analysis and integration of Subject Matter Experts' (SME) feedback within a systematic migration workflow. The methodology achieves significant reductions in Syntax Error Rates (SER), enhances feature alignment throughout migration iterations, and leverages dataset sampling to ensure continual improvement. By embedding GAI into the migration process, the framework facilitates precise feature mapping, semi-automated error resolution, and data-driven optimisation loops, improving workflow efficiency.**

## I. Introduction

Migrating the codebase of a large database management system (DBMS) from Oracle to PostgreSQL is a complex and multifaceted process that goes far beyond simple syntax conversion. This is a complex, interdisciplinary task that requires ensuring the semantic integrity of business processes, minimising architectural risks, ensuring business continuity, and achieving similar or even better performance across different architectures of these database management systems. This problem is complicated by several important aspects. First, the structure of SQL code libraries is often highly dependent, making it very difficult to automate the analysis and transformation of large code libraries containing hundreds of thousands of files. Second, the SQL dialects Oracle PL/SQL and PostgreSQL PL/pgSQL differ significantly in syntax, structure, data types and performance configuration, making it difficult to ensure compatibility of data types, schemas and logic [1-2].

Tools like AWS Schema Conversion Tool [3] and Ora2Pg [2] currently offer the ability to convert underlying database objects using schema conversion, but their capabilities for deeper contextual analysis and automation of key elements such as triggers, process packages, dynamic SQL and exceptions are limited. Oracle and PostgreSQL have significant differences in query optimisers, indexing strategies, cursor behaviour and SQL query processing mechanisms [1]. This means that without a more integrated, adaptive and iterative approach, the "direct migration" method cannot be applied to the migration of large-scale code libraries.

In recent years, significant progress has been made in transforming code libraries using large language models (LLMs) and generative artificial intelligence (GAI). Open source models such as Code Llama [4] and StarCoder2 [5] have demonstrated significant improvements in the accuracy of syntactic translation and code synthesis. However, as numerous empirical studies have shown, LLMs do not always provide perfect quality during code library migration, allowing systematic errors such as semantic drift, API call errors, and inconsistent control structure mappings [6]. These observations point to the need to use subject-oriented models and integrate LLM into multi-stage pipeline processes that include static analysis, semi-automated error handling and iterative verification.

Methodological frameworks play a special role in improving migration performance. An important task is to maintain semantic compatibility and minimise code loss, especially in complex cases such as converting Oracle NUMBER to PostgreSQL NUMERIC or PL/SQL cursors to PL/pgSQL structures [1]. The current problem is the lack of a unified scientific basis that would allow for taking the global context into account during the development, model tuning and error handling phases of inference of complex migrations. In addition, there is a demand to develop effective metrics and methods for analysing losses, their causes, and their impact on the final product.

A fundamental methodological challenge remains to find the optimal balance between fine-tuning and rapid design. Fine-tuning allows GAI to be tailored to the task using domain data, but this approach requires more resources and access to large datasets [7]. Conversely, while high-quality prompt engineering is less expensive, it may not be enough to solve all the problems associated with semantic differences between Oracle and PostgreSQL.

This article addresses all of the above issues and lays the foundation for creating an integrated iterative framework designed to automate the migration of codebases from Oracle to PostgreSQL. The proposed method integrates the various generative artificial intelligence methods, including LLM fine-tuning mechanisms, signal engineering, the use of Retrieval Augmented Generation (RAG), static analysis and formal metrics for assessing translation quality, which will help create an improved and manageable translation process.

Based on the conducted scientific analysis, it is clear that migrating the SQL codebase from Oracle to PostgreSQL is a challenging task, given the significant differences between Oracle and PostgreSQL. Effective automation of this process requires the use of advanced context-handling techniques suitable for LLM to achieve syntactic and semantic compatibility, as well as iteratively driven error correction and optimisation processes. For a long time, the scientific community lacked a universally applicable scientific basis to ensure the correctness and effectiveness of large-scale migration.

Therefore, among the scientific problems of migration, the following should be considered:

- Develop methods for analysing large-scale context-aware code libraries containing over 100,000 files.
- Identify conditions for the optimal choice between fine-tuning and rapid design, as well as methods for combining them.
- Create an iterative architecture for controlled quality improvement, error prediction and quality maintenance at every stage of the migration process.
- Quantify functional semantic losses and integrate them into a semi-controlled error correction process.
- Building a methodological foundation that includes the integration of neural network models, static analysis and the RAG approach.

Thus, we need to develop a hybrid framework capable of providing automated, context-sensitive, iterative, and reliable migration of SQL codebase from Oracle to PostgreSQL by implementing innovative algorithms and Generative AI techniques and formalising processes.

## II. METHODOLOGY

### *A. Two-stage fine-tuning rationale*

The LLM preparation approach (Fig. 1) represents a tightly coupled iterative framework.

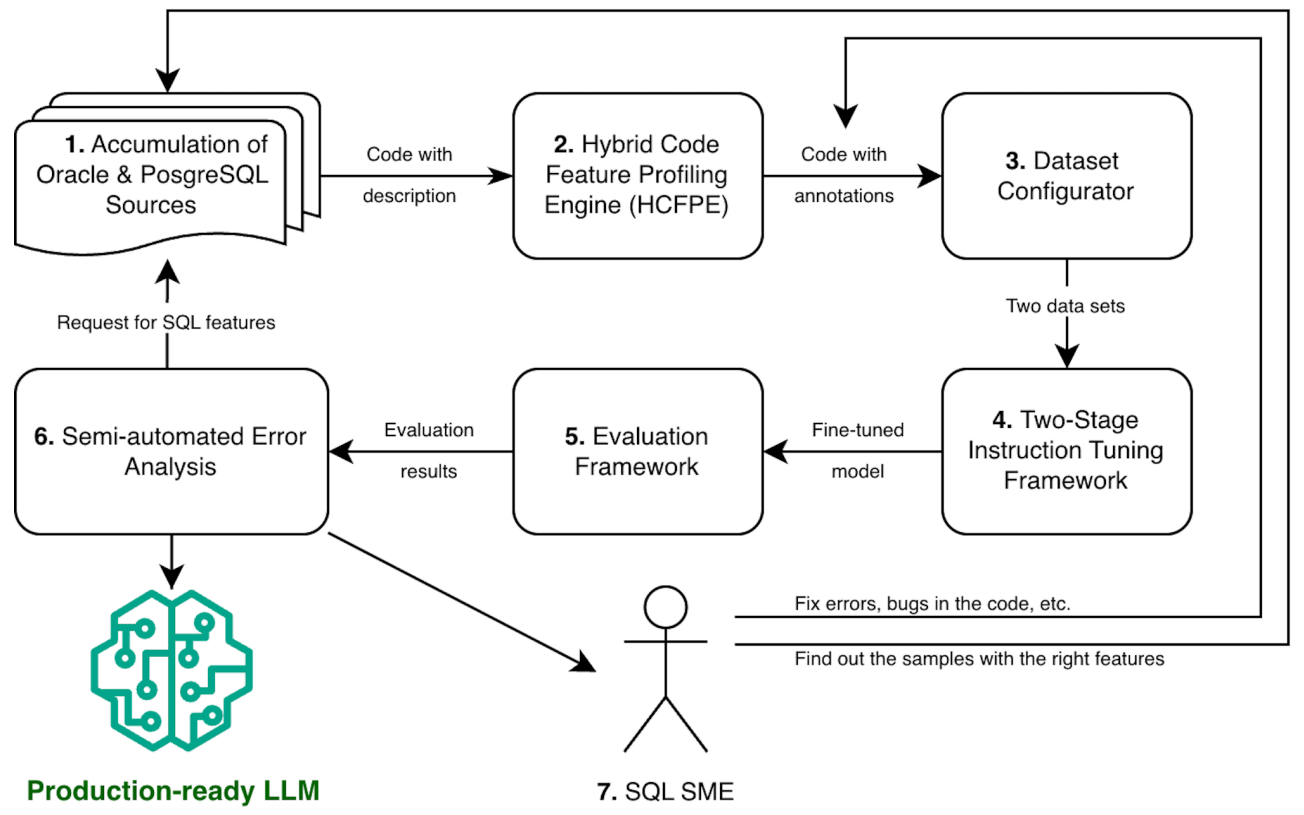

Fig. 1. Production-ready LLM preparation pipeline.

The framework allows to build a domain-specific, large-scale LLM capable of performing accurate, context-aware migrations from Oracle to PostgreSQL. This approach does not treat fine-tuning as a single, holistic learning event, but rather organises it into a series of interdependent computational and human-computer interaction phases. It collectively ensures syntactic accuracy, semantic preservation, and high-coverage feature transformation.

The architecture of LLM preparation pipeline reflects the empirical finding that reliable database migration cannot be achieved solely through direct prompt. Instead, it requires an iterative improvement cycle in which the model learns SQL syntax and observes representative patterns. This receives carefully selected feedback and iteratively reduces the number of typical error patterns in the multi-dialect codebase.

### *B. Role of HCFPE in dataset construction*

The process begins by collecting various Oracle and PostgreSQL data sources, which form the initial foundation for subsequent dataset development. These sources include stored procedures, triggers, package bodies, SQL*Plus scripts, database management utilities, and RMAN fragments, each with distinct syntactic and semantic features. Because these artefacts exhibit significant internal diversity, static reading alone is insufficient. They must first be processed by the Hybrid Code Feature Profiling Engine (HCFPE), which annotates the corpus using SQL feature metadata. This stage extracts syntax markers, flow structures, control flow patterns and dialect-specific operators, effectively converting unstructured code into a structured representation suitable for configuring deterministic datasets. Without this feature-based processing, subsequent stages will be unable to assemble a consistent training set that can capture the distribution patterns of the problem domain.

Then, the annotated corpus is being processed by a dataset configurator to create two independent but complementary datasets supporting a two-step fine-tuning strategy. The first dataset captures syntactic-level correspondences, where the mappings are implemented using natural language descriptions. This example follows the pattern of “<Code *(Oracle / PostgreSQL)> : <Description>”*. It allows the model to abstractly understand SQL structure, version differences and syntactic equivalence without direct conversion. By learning to express code as natural language descriptions and reconstructing SQL from these descriptions, the model learns structural patterns that are crucial for downstream migration tasks. In contrast, the second dataset contains direct transformation pairs, “*<Oracle code> : <PostgreSQL code>”,* without an intermediate description layer. These transformation pairs form the basis for the second-stage fine-tuning, where the model shifts from syntactic understanding to understanding the transformation behaviour of the execution dialect.

These two datasets were used to establish a two-stage instruction tuning framework. In the first stage, the model was optimised to learn the syntax of a specific domain. The goal of this stage was not transformation, but alignment: the model learned the syntax of Oracle and PostgreSQL, nuances at the dialect level, type signatures, scope rules, exception propagation semantics and version-specific keyword behaviour. Only after this structural competence is established does the second stage of fine-tuning occur. This is what allows for the actual conversion between dialects. During this stage, the model learns to preserve semantics while transforming procedural blocks, dynamic SQL, exception structures, cursor logic or proprietary Oracle constructs into appropriate PostgreSQL equivalents. The two-stage approach ensures that errors caused by insufficient syntactic foundations do not contaminate the transformation stage, thereby improving convergence and generalisation capabilities.

### C. *Human-in-the-loop error correction*

After two-stage optimisation, the model output is fed into an evaluation system that applies a series of quantitative and structural validation procedures. The chosen metrics, such as completeness, BLEU and CHRF, are used to assess the apparent accuracy of the grammar. At the same time, the SQL code checker, static analyser and PL/pgSQL verifier are used to verify the correctness of the executable files.

However, for domains where correctness is closely related to deep semantic preservation, numerical metrics alone are insufficient. Thus, the evaluation integrates structure checking, feature coverage analysis, and difference identification, which is consistent with SQL feature classification methods extracted in the early stages of the pipeline. The evaluation results and failure cases are then compiled into a comprehensive error analysis report. The semi-automatic error analysis module interprets errors and othe the issues down into groups of coverage:

- syntax;
- semantics;
- structural;
- functional.

For Oracle-specific or version-specific features that are not yet fully covered, the system requests additional samples from the knowledge base or from experts. This step transforms error analysis into a dataset optimisation mechanism, enabling targeted resampling and the synthesis of new paradigms to address system performance issues.

SMEs selectively participate in this phase of the work. They do not manually review every sample but only intervene when a sample is labelled as ambiguous, structurally complex or semantically incompatible. Expert's feedback is used to expand or modify training data, improve annotations or adjust query templates. Thus forming a feedback loop for human-machine collaboration without incurring unnecessary overhead. The experts' recommendations will then be fed back into the data accumulation and dataset adjustment phases, providing a better model for the next round of fine-tuning.

### D. *Closed-loop iteration logic*

Ultimately, through iterative cycles of syntax tweaking, transformation tweaking, evaluation and error-based improvements, the system produces a ready-to-use LLM capable of performing high-precision, context-aware migrations from Oracle to PostgreSQL. The closed-loop design of this pipeline ensures that the number of error patterns observed in previous loops is gradually reduced with each iteration, while the dual-dataset design ensures the coordinated development of syntactic understanding and transformation capabilities. Therefore, the final model is not only a translator, but also a topic-oriented logical reasoning mechanism equipped with a structured internal representation of the SQL dialect, its feature distribution, and semantic correspondences.

Integrating different types of knowledge into a single knowledge base simplifies search logic and reduces operational complexity, as only a single FAISS index needs to be maintained, updated, and queried. Search latency is reduced, and the risk of inconsistencies or out-of-sync issues across multiple repositories is eliminated.

### E. *Knowledge Base Architecture*

The system supports two architecture strategies for building the knowledge base used in RAG during Oracle to PostgreSQL migration. Strategy A employs a hybrid knowledge base architecture, storing Oracle code blocks, PostgreSQL archives and transformation rules written by domain experts in three separate vector FAISS databases. This process, as shown in Fig. 2, first involves chunking, embedding and storing the Oracle program code and PostgreSQL documentation. These independent semantic stores enable LLM to extract heterogeneous contextual clues during migration: structural patterns from Oracle code, authoritative syntax rules from PostgreSQL archives and deterministic transformation logic derived from SME analysis. This multi-repository design improves search reliability by allowing the system to combine application context, dialect rules, and error correction logic, even if a single source cannot cover the data.

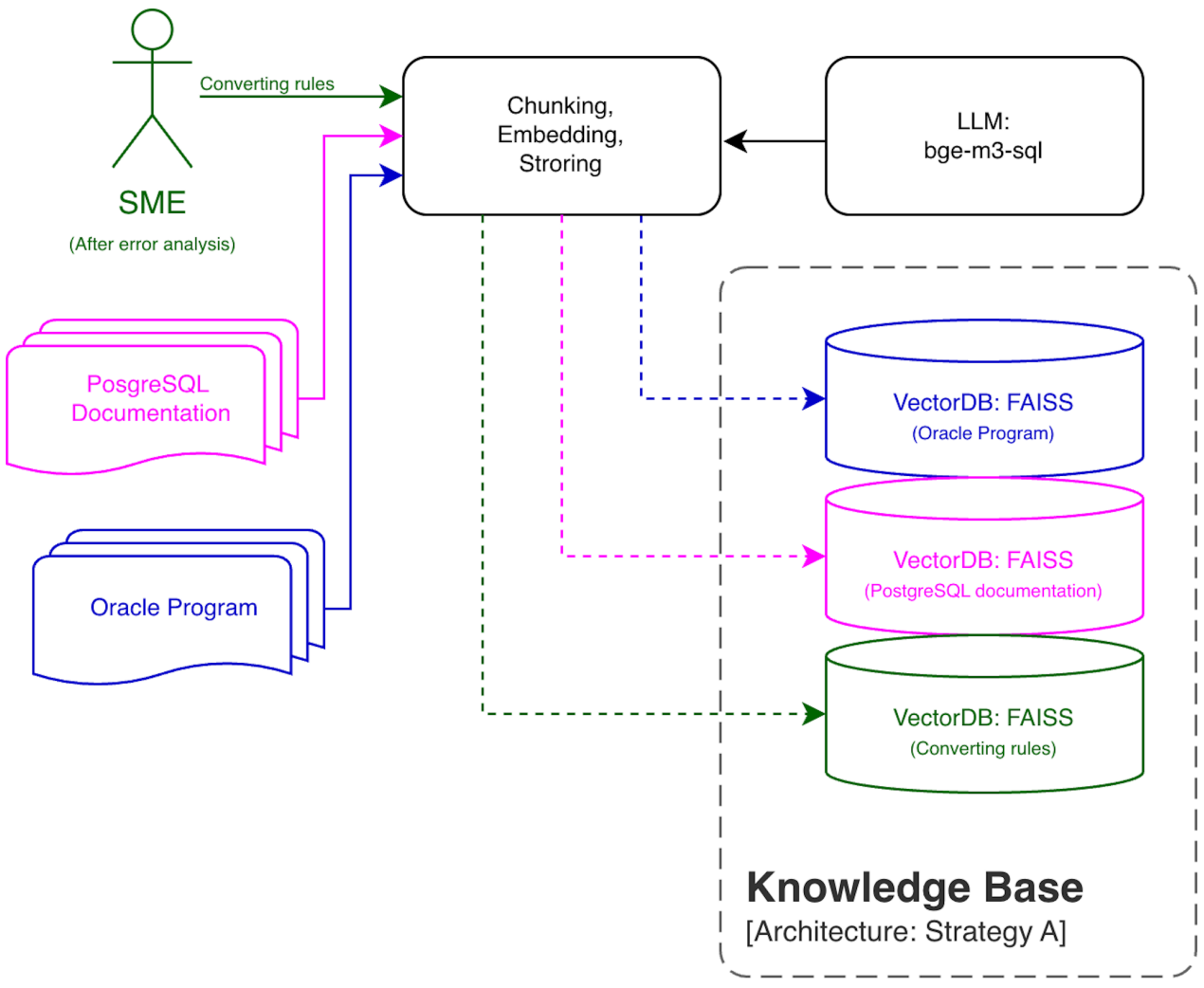

Fig. 2. Knowledge Base Architecture: Strategy A.

However, unification also eliminates the structural separation of different types of knowledge, which is a key advantage of Strategy A. Oracle code, PostgreSQL reference archives, and rule-based transformations are intertwined within a single vector space. A unified vector database is feasible, but it sacrifices the fine-grained control and high contextual accuracy offered by Strategy A.

As shown in Fig. 3, Strategy B uses a unified vector database consisting exclusively of fragmented embeddings from Dataset 2, which contains direct "Oracle-PostgreSQL" code pairs. This approach compresses the search process into a simplified architecture where transformation paradigms dominate the knowledge space, thereby reducing operational complexity and accelerating the search process.

This architecture is specifically designed to test whether a basic, unadjusted LLM can perform accurate SQL

transformations based solely on the obtained examples, without updating domain-specific parameters. Strategy B relies entirely on paired examples, and its effectiveness is closely related to the completeness of the dataset: missing or underrepresented Oracle features cannot be compensated for by documentation or rule-based instructions. Therefore, Strategy B works well when the dataset can cover the migration scope well, but it may cause problems for rare PL/SQL structures, management scripts, or non-standard Oracle extensions.

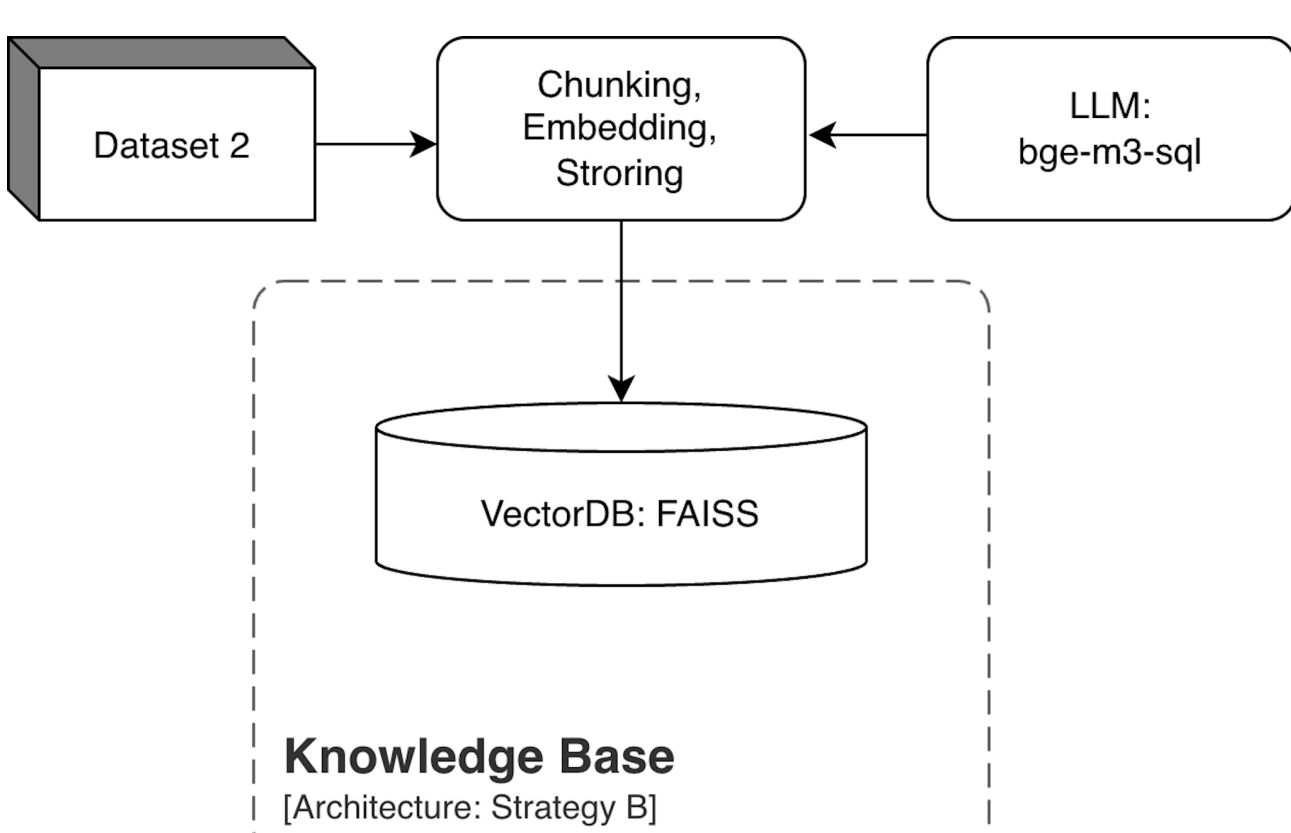

Fig. 3. Knowledge Base Architecture: Strategy B.

Strategy A provides robustness and high flexibility for complex functions, while Strategy B offers simplicity, fast data acquisition, and high performance when high-quality paired datasets are available.

### *F. RAG and Prompting*

Both Strategy A and Strategy B use RAG to improve the accuracy and context relevance of Oracle to PostgreSQL migrations. In both cases, the RAG process starts with the input Oracle code chunks and transforms them into embedding vectors using the same embedding scheme constructed from the dataset. Therefore, the search principle of the two strategies is the same: the system determines the closest match in the vector space. It uses these matches to enrich the contextual understanding of the language learning model (LLM) before generation. Then, the vector is used to query the knowledge base, and the most semantically similar entries are extracted based on the similarity of the nearest neighbours.

The difference lies entirely in the knowledge returned. In Strategy A, the search results generate heterogeneous triples: Oracle context fragments, PostgreSQL document extraction and transformation rules derived from SME, with each triple coming from a separate vector store. In Strategy B, the search returns only samples of “Oracle-PostgreSQL” transformation pairs obtained from a single unified vector database built on dataset 2. Therefore, Strategy A provides recommendations based on multiple sources, rules and documents. In contrast, Strategy B provides recommendations entirely based on examples and templates, aiming to test how well the underlying LLM can generalise without fine-tuning.

After the search is complete, the system inserts the retrieved content into the corresponding Strategy A or Strategy B suggestion template to optimize the recommendations. Then, LLM generates PostgreSQL transformations for the current data block. Ultimately, all generated blocks are aggregated and reassembled into a coherent PostgreSQL script, thus preserving the structure of the original Oracle program.

We used different experimental configurations, changing the strategy, prompts, model and number of retrieved neighbours, to evaluate how retrieval combinations affect translation quality and robustness. The appropriate prompt for Strategy A is shown in Fig. 4, and for Strategy B is shown in Fig. 5.

```
PROMPT = """
You are converting an Oracle SQL/PLSQL chunk to PostgreSQL.

Context:
• Oracle reference examples (global context): {ORACLE_CONTEXT}
• PostgreSQL documentation excerpts (syntax and semantics): {POSTGRES_DOCS}
• SME converting rules (must override examples when applicable): {CONVERTING_RULES}

Current chunk to convert:
{CURRENT_CHUNK}

Task:
Produce the PostgreSQL equivalent of the current chunk.
Follow these principles:
1. Apply converting rules with highest priority.
2. Use PostgreSQL documentation to ensure canonical and version-correct syntax.
3. Use Oracle context only to understand intent, not for direct copying.
4. Do NOT hallucinate missing constructs; if ambiguity exists, choose the safest syntactic equivalent.
5. Output only valid PostgreSQL code.
"""
```

Fig. 4. Structure of a prompt for Strategy A.

The provided prompt is a prompt template designed for Strategy A, where the LLM receives three heterogeneous knowledge sources retrieved from separate vector databases. The prompt instructs the model to synthesise these three sources while performing a context-aware conversion of the currently processed Oracle chunk.

```
PROMPT = """
You are converting an Oracle SQL/PLSQL chunk to PostgreSQL.

You are provided with semantically similar Oracle→PostgreSQL conversion examples
retrieved from the knowledge base. Use them as guidance for style, structure,
and feature mapping.

Retrieved examples (Oracle → PostgreSQL):
{RETRIEVED_EXAMPLES}

Current Oracle chunk to convert:
{CURRENT_CHUNK}

Task:
Generate the PostgreSQL equivalent of the current chunk using ONLY the patterns
and mappings observable in the retrieved examples.

Guidelines:
1. Learn from the examples: follow their mapping patterns, syntax choices,
   type conversions, and procedural rewrites.
2. Do NOT hallucinate undocumented constructs; if the pattern is not visible
   in the examples, produce the safest syntactic PostgreSQL alternative.
3. Preserve semantics: maintain cursor logic, exception behaviour, and SQL flow
   consistent with the Oracle intent.
4. Output only PostgreSQL code without explanations, comments, or additional text.
"""
```

Fig. 5. Structure of a prompt for Strategy B.

Conceptually, it operationalises the core principle of Strategy A—triangulation: the LLM must interpret the Oracle code, apply deterministic rules, and align the output with PostgreSQL syntax and semantics as documented in authoritative reference material. Prompt for Strategy B is the instruction template, where the knowledge base contains only example-driven “Oracle-PostgreSQL” conversion pairs. Thus, the prompt operationalises the core principle of Strategy B, which is based on example-driven contextual grounding. In this approach, conversion behaviour arises from similarity-guided retrieval rather than from explicit rule enforcement.

### *G. Dataset Preparation*

Dataset evaluation is a critical step in the Oracle to PostgreSQL migration process, allowing for systematic optimisation of training data based on actual performance. Through feedback analysis and quantification, this stage can identify structural and feature defects in the dataset.

The dataset was stratified by file size: small, medium and large. It helps to reflect realistic codebase heterogeneity and to determine how model performance scales with increasing structural and semantic complexity. This stratification allows for controlled analysis of size-dependent error patterns, sensitivity to chunking and model capability boundaries. Collectively, these experiments provide an empirical foundation for selecting the most effective configuration for enterprise-scale SQL migration.

To better understand the structure of the actual Oracle codebase, we analyzed the distribution of Oracle functions across all scripts, as shown in Fig. 6. This research is necessary because a single script often contains several different types of functionality, such as a combination of PL/SQL blocks and SQL*Plus commands, or a combination of static SQL, dynamic SQL, and control structures.

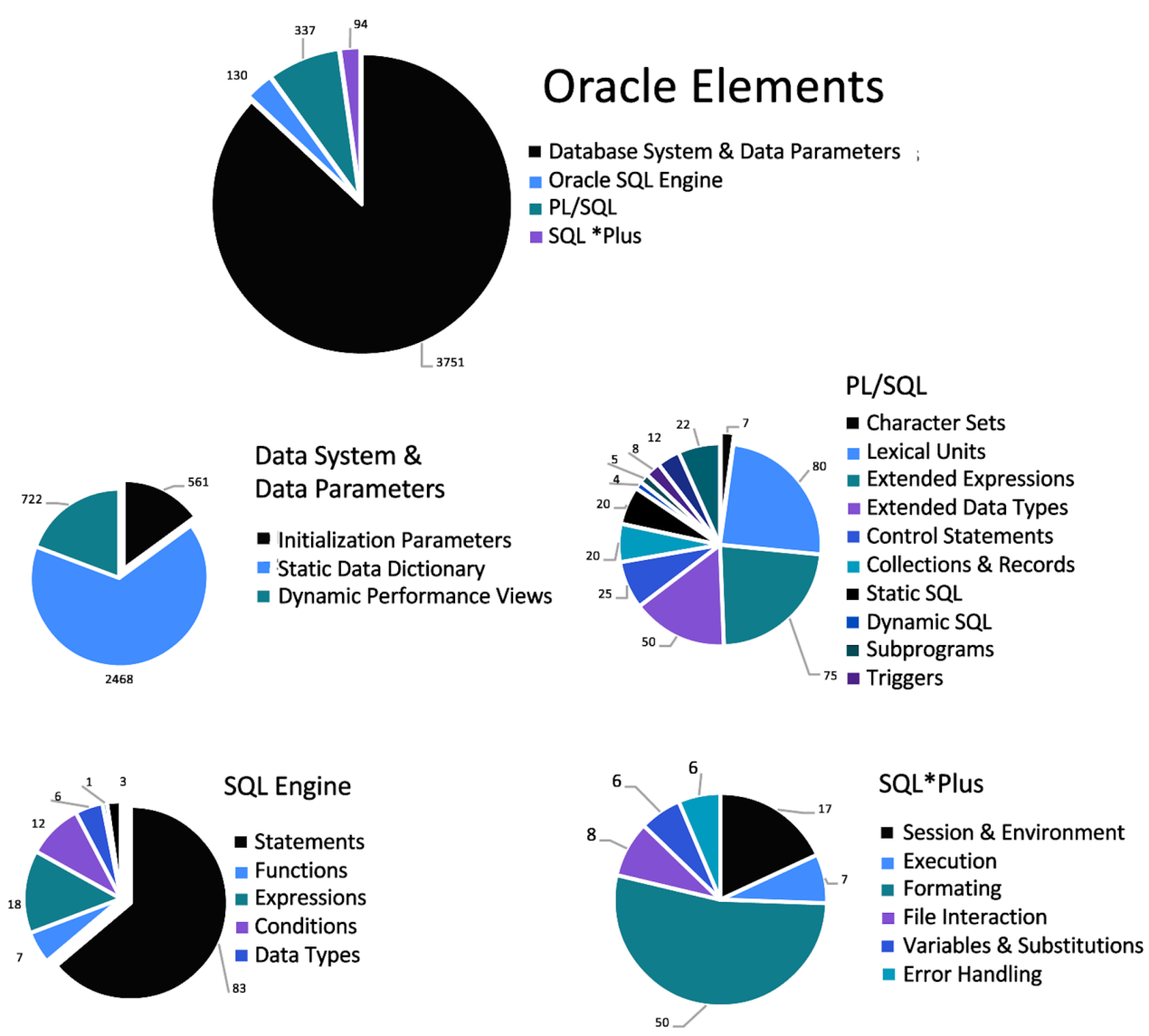

Fig. 6. Oracle Samples Distribution by Feature in Datasets.

The coexistence of these characteristics complicates the migration process because different Oracle features impose different syntactic and semantic constraints. By quantifying the distribution of these features globally (Fig. 7) and in a single scenario, we can accurately understand feature density, overlap and complexity.

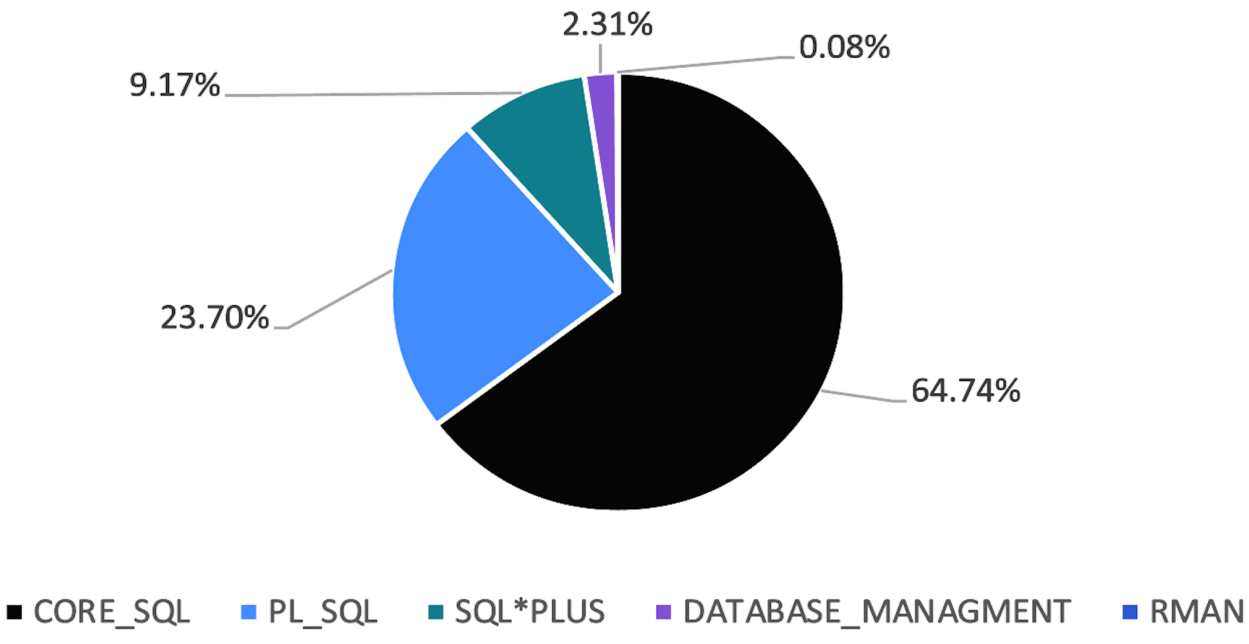

Fig. 7. Generalised feature distribution in a typical Oracle file.

This analysis directly guided dataset construction, partitioning strategies, feature-based scoring and prioritisation of complex feature groups, such as RMAN and SQL*Plus, ultimately enabling finer-grained tuning and more robust Oracle to PostgreSQL conversion.

Based on our HCFPE (Fig. 1), we obtained a detailed distribution of Oracle features across all analysis scenarios. HCFPE performs structure checks and semantic classification at the keyword level, thereby breaking down each script into its constituent Oracle subsystems, including SQL engine constructs, PL/SQL elements, SQL*Plus commands and data system parameters. Therefore, the engine generates a quantisation graph describing the frequency of occurrence of each feature category in the codebase. Table 1 summarises the results of this analysis process, outlining the global distribution of Oracle features.

TABLE I. Oracle Feature Distribution in Datasets

| N | Feature Name | Train, % | Test, % |
|---|---|---|---|
| 1 | CORE_SQL | 57,94 | 57,55 |
| 2 | PL_SQL | 27,21 | 24,50 |
| 3 | SQL_PLUS | 13,72 | 16,63 |
| 4 | DATABASE_MANAGEMENT | 1,09 | 1,25 |
| 5 | RMAN | 0,04 | 0,06 |

The HCFPE decomposes Oracle application features into detailed keyword-level representations suitable for both training and test sets. This distribution is crucial for understanding feature density, identifying mixed feature scenarios, and guiding subsequent steps:

- dataset sampling;
- block allocation strategies;
- prioritisation of model fine-tuning and evaluation.

To supplement the analysis of the features obtained from Oracle, we also examined the distribution of PostgreSQL features in all generated and used PostgreSQL scripts, as shown in Fig. 8.

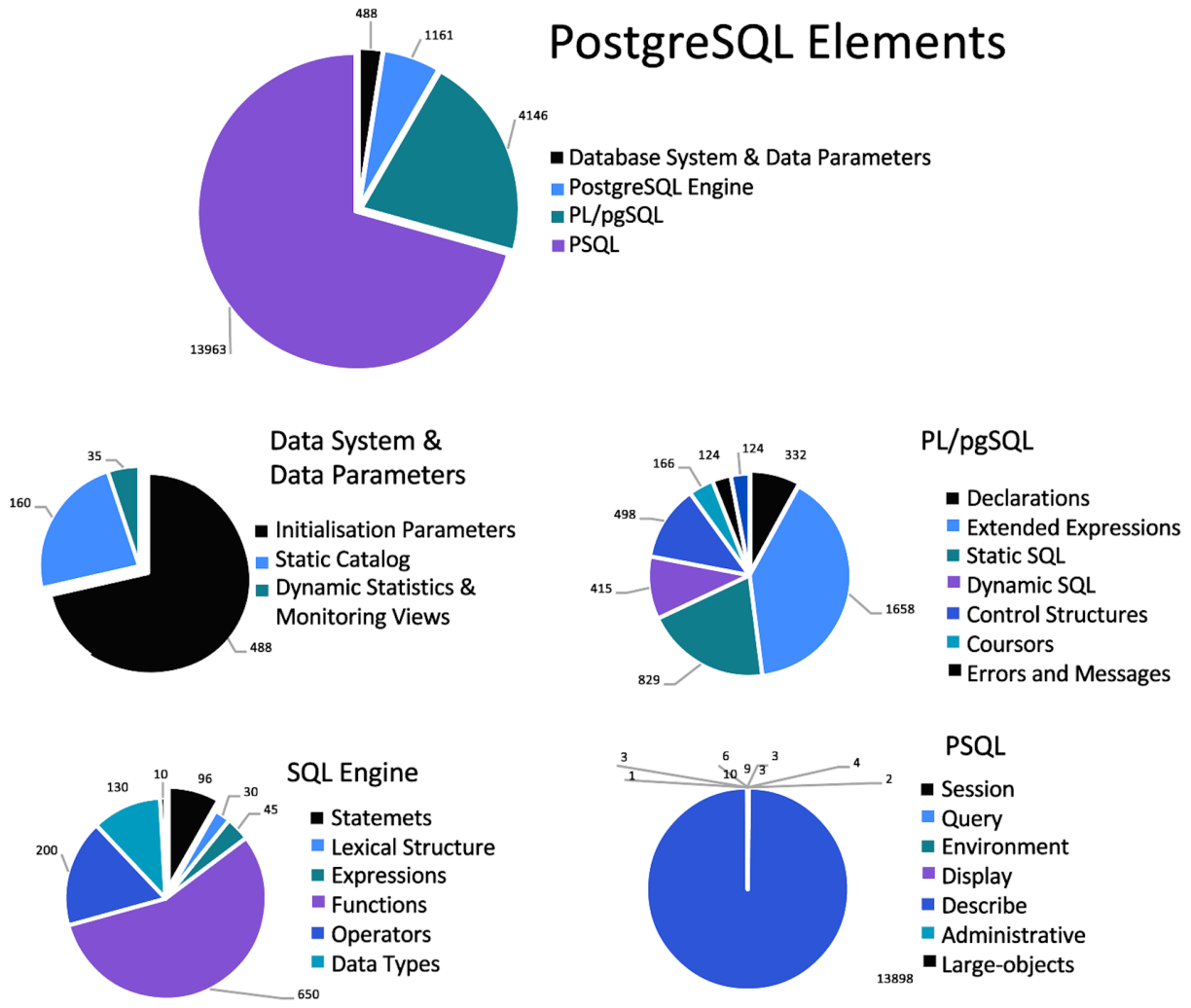

Fig. 8. PostgreSQL Samples Distribution by Feature in Datasets.

This research is necessary because a single PostgreSQL script typically contains several interactive subsystems, such as a combination of PL/pgSQL flow structures and PSQL script commands, or a combination of SQL engine expressions and directories. Understanding these mixed feature patterns is crucial for assessing the complexity of transfer and identifying semantic biases that arise during the transformation process.

Furthermore, PostgreSQL feature distribution analysis helps build a predictive engine for further validation. By examining the expected distribution of PostgreSQL features corresponding to specific Oracle feature combinations, the system can predict which feature combinations should be included in the correctly generated code. During the evaluation process, the predicted feature distribution can be compared with the actual distribution in the PostgreSQL output generated by the LLM. Significant deviations between the expected and generated distributions may indicate structural inconsistencies, a lack of transformation, or the presence of spurious structures. This feature-based verification method adds a layer of quality control on top of standard grammar metrics, enabling more accurate detection of conversion anomalies in complex porting scenarios.

To conduct a systematic comparison of the datasets, we first summarised the PostgreSQL feature distribution obtained by the mixed code feature analysis engine and visualised it in Fig. 9.

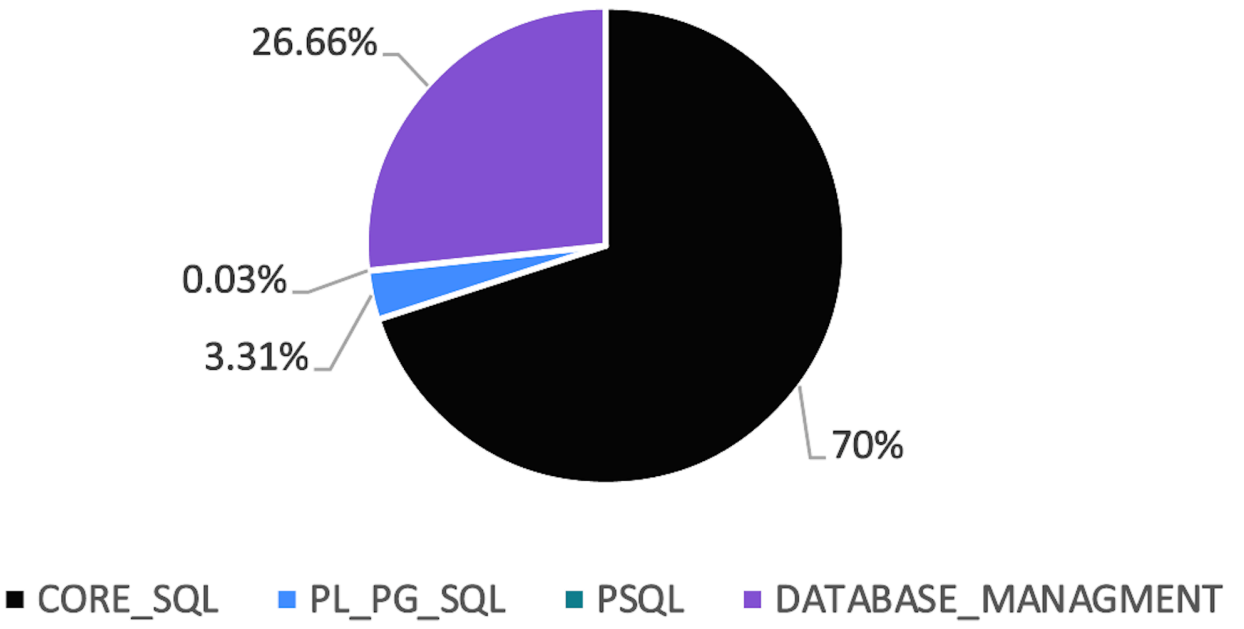

Fig. 9. PostgreSQL Samples Distribution by Feature in Datasets.

These distributions summarise the relative popularity of PostgreSQL subsystems, such as SQL engine constructors, PL/pgSQL process elements, PSQL commands and system-level parameters, across all analysed scripts. After obtaining these global proportions, we converted the feature counts into normalised percentage values and reported these values in Table 2, which shows the proportion of each PostgreSQL feature category in the training and test sets.

TABLE II. PostgreSQL Feature Distribution in Datasets

| N | Feature Name | Train, % | Test, % |
|---|---|---|---|
| 1 | CORE_SQL | 62,83 | 61,89 |
| 2 | PL_PG_SQL | 4,40 | 2,94 |
| 3 | DATABASE_MANAGEMENT | 32,66 | 35,16 |
| 4 | PSQL | 0,11 | 0,02 |

This tabular representation allows us to quantify the feature balance between different datasets, identify underrepresented or overrepresented structures in PostgreSQL, and ensure that model estimates are not biased by feature frequency bias. Normalising the feature distribution also lays the foundation for comparing the expected and generated features used in subsequent validation and error analysis stages.

This structured iterative sampling strategy ensures that subsequent fine-tuning cycles obtain optimally balanced and impactful training data. Therefore, the transmission quality is improved during the iteration process, the error rate gradually decreases, and the stability of feature transformation is also improved.

### H. Experimental System Configurations

These experiments were conducted to systematically evaluate how different LLMs and different RAG configurations perform across the diverse conditions encountered in real-world Oracle-to-PostgreSQL migration. By testing base and fine-tuned variants of Qwen, multiple GPT families and the traditional Ora2PG tool, the study aims to quantify the relative contribution of model size, fine-tuning, and architecture choice to translation quality, robustness, and error behaviour. Examining models across four pipelines: direct conversion (Fig. 10), history-aware conversion (option for entire program migration) (Fig. 11), RAG mode with Strategy A/B (Fig. 12), which enables isolation of the specific benefits introduced by sequential context accumulation and RAG.

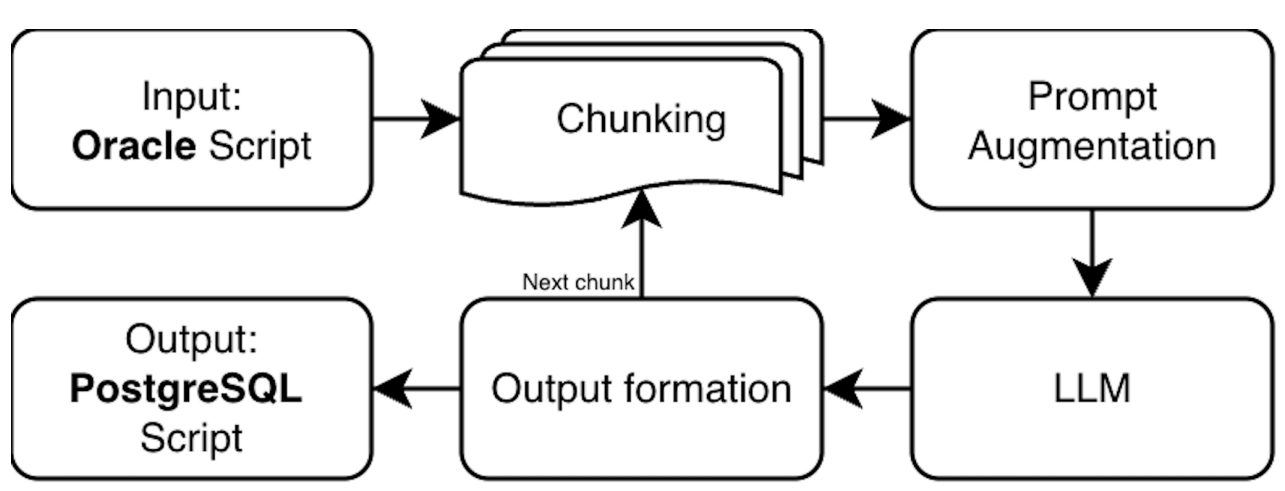

Fig. 10. Component-level of the conversion pipeline.

The conversion process directly performs the Oracle to PostgreSQL conversion without retrieving any content from the knowledge base. The Oracle input script is segmented into manageable and semantically coherent blocks. Each block is processed by a prompt enhancer, which generates model-specific instructions, after which the LLM builds the corresponding PostgreSQL fragment. The source module is then sequentially assembled with all translated code chunks to reconstruct the final PostgreSQL script. This process evaluates the translation capabilities of LLM without external context support.

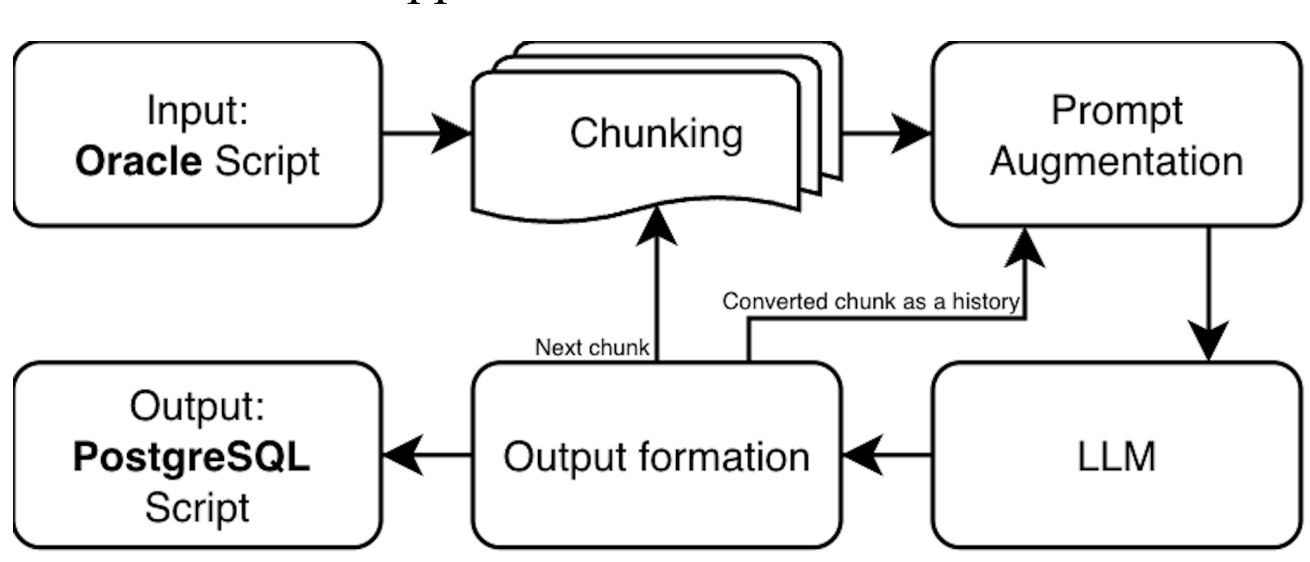

Fig. 11. Component-level of the history-aware conversion pipeline.

The history-aware process extends the capabilities of direct conversion by incorporating pre-generated fragments as contextual history information. Oracle scripts are divided into multiple chunks, each of which is processed sequentially. Before broadcasting, the prompt improvement steps will input the current chunk and the accumulated broadcast history, enabling LLM to maintain state consistency across chunks. No search or knowledge base was used - all contextual information comes from the results of previous transformations.

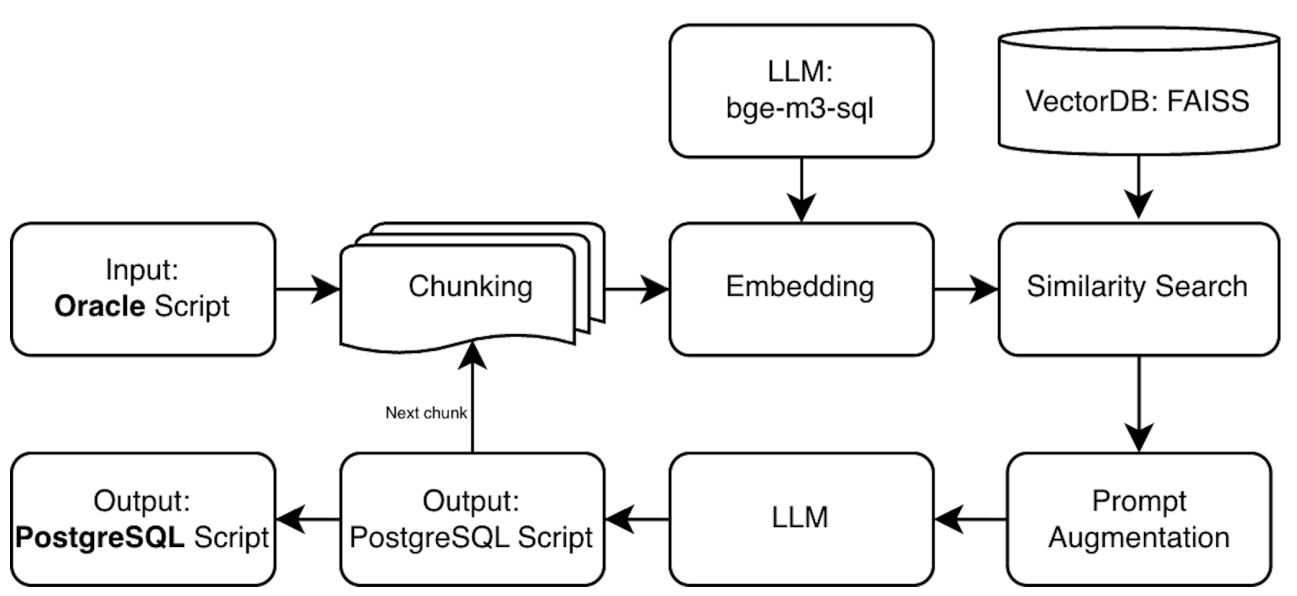

Fig. 12. Component-level of RAG mode with Strategy A/B pipeline.

In the RAG configuration (Fig. 12), the Oracle to PostgreSQL conversion process integrates the ability to retrieve information from external knowledge bases to improve contextual relevance and reduce conversion errors. First, the process segments the input Oracle script into semantically consistent chunks. Each code chunk is embedded using the same embedding model as during the knowledge base construction process. The resulting embedding vectors are used to perform similarity searches in the FAISS vector database. For strategy A, the heterogeneous components, Oracle context fragments, PostgreSQL file extraction and SME transformation rules. It is worth noting that Strategy A requires storing the entire Oracle application corpus in a knowledge base to support example-based contextual searches. In contrast, Strategy B only includes Oracle-PostgreSQL pairwise transformation examples obtained from dataset 2.

The acquired information is integrated during the prompt augmentation stage and combined with the current Oracle data chunks and policy-specific prompt. The enhanced prompts are processed by LLM to produce PostgreSQL transformations. Then, in the output generation stage, all transformed code chunks are assembled into a complete PostgreSQL script. This supports the RAG process to assess how retrieval quality and knowledge base structure affect translation consistency and feature coverage.

To ensure the search components function reliably before the full migration experiment, we created a gold standard dataset specifically for testing and validating the search engine. This dataset is designed to cover a variety of real-world search scenarios that arise during the migration from Oracle to PostgreSQL. This includes cases where a given Oracle code chunk has a clear contextual match with the corresponding PostgreSQL instance in the knowledge base, as well as cases where the code chunk is intentionally missing any contextual counterpart, thus enabling us to test the robustness and fallback behavior of the search. In addition, other intermediate scenarios have been added to detect partial matches, mixed feature chunks, ambiguous contexts and patterns that are syntactically similar but semantically different.

Using this valuable dataset, we systematically evaluated the search accuracy, ranking stability, and sensitivity to context sparsity of strategies A and B. We will only conduct a full experimental evaluation of the RAG-based end-to-end pipeline after confirming that the search engine can correctly identify relevant neighbors in all use case categories.

### I. Evaluation Framework and Code Validation Approach

Our evaluation framework provides an independent and flexible mechanism for assessing the quality of Oracle to PostgreSQL migrations. It supports multiple evaluation criteria and experimental scenarios, enabling comprehensive analysis of model behavior without actually performing the conversion. This framework takes the original Oracle script and the resulting PostgreSQL output, along with the corresponding real PostgreSQL script, and calculates several types of metrics.

To quantify the accuracy of the translation, we used Recall, BLEU and ChrF metrics, which measure the lexical, syntactic, and structural similarity between the translated text and the source text, respectively. In addition, we calculated the Normalized Syntax Error Rate (NSER) and Normalized Warning (NSWAR), which reflect the correctness of the execution level. We used SQL code verification to detect syntax violations and conducted preliminary logic verification experiments, such as generating database schemas and attempting partial script execution, although full execution level verification was not the primary goal of this study.

In addition to translation metrics, the scoring system also evaluates the accuracy of segments to ensure that structural segmentation is preserved, and assesses search quality by checking whether the RAG subsystem (if used) returns semantically relevant neighbors. Natural Language Processing (NLP) metrics ensure the output fully meets expectations, while syntax checking and warnings ensure the code is executable. We used SQL code checking tools for syntax verification to detect structural errors. We also executed some preliminary experiments on logic verification, namely, generating appropriate database schemas and attempting partial execution of translated scripts. However, complete logic execution verification is not the primary goal of this study.

### J. Dataset estimation methodology

To enable continuous improvement of the fine-tuned LLM, we developed a dataset-estimation mechanism that identifies which SQL features require additional training samples in the next iteration. This mechanism combines information from four sources: the existing dataset, the SQL feature taxonomy, the results of error analysis, and per-feature quality metrics. The outcome is a feature-level performance gap, $GAP_{Feature}$, which represents how far the model is from the desired target accuracy for each feature category and determines how many new samples must be collected.

A central component of this methodology is $GAP_{Dict}$, which measures the degree to which each SQL feature is represented in the dataset. Because LLM performance strongly depends on training distribution, underrepresented features lead to systematic weaknesses in conversion quality. $GAP_{Dict}$ is computed by normalizing the training sample count of each feature relative to the most frequent feature in the dataset:

$$GAP_{Dict}(f) = 1 - \frac{Count_{train}(f)}{max_{all\,features}(Count_{train})}. \qquad (1)$$

This formalisation ensures that well-represented features produce values near zero, indicating that the dataset already provides sufficient coverage. Conversely, rare features, such as DB_MAN or RMAN, produce values near one, revealing that the model lacks exposure to these constructs and requires targeted sampling. In practice, this measure directly prioritizes which SQL patterns SMEs need to annotate next.

Formula (2) calculates weighted raw quality of feature conversation by LLM:

$$Q_{Raw} = \omega_R R + \ ... \ + \omega_{SER}(1 - SER),$$

where all metrics are converted to a single score in [0,1] and $(1 - SER)$ because low $SER$ is good.

For normalising we should:

$$Q_{Norm} = \frac{Q_{Raw}}{\omega_R + \omega_B + \omega_C + \omega_{Agg} + \omega_{ser}}.$$

So, $GAP_{Quality}$ is:

$$GAP_{Quality} = 1 - Q_{Norm},$$

where:

- $GAP_{Quality} \approx 0$: feature already very good;
- $GAP_{Quality} \approx 1$: feature very bad, needs much improvement.

The overall $GAP_{Feature}$ is measured in "%" and then computed as:

$$x = (1 + \beta^2)(1 - GAP_{Quality})(1 - GAP_{Dict})$$

$$GAP_{Feature} = \left(1 - \frac{1}{2-x}\right) \times 100\% \qquad (5)$$

where $GAP_{Quality}$ aggregates model performance using weighted values of Recall, BLEU, ChrF, aggregated score, and syntax-error rate. The weighting vector β control the importance of translation accuracy relative to dataset sufficiency.

This methodology supports two key capabilities:

- Targeted improvement of weak features. When a feature shows high $GAP_{Feature}$, the framework identifies which training examples must be added or rebalanced to close the performance deficit in the next fine-tuning cycle.
- Automatic support for new SQL features. Newly introduced features initially yield $GAP_{Feature}$ near 100%, causing the system to automatically request the required samples, enabling seamless expansion of LLM capabilities across SQL dialects and toolsets.

The pipeline aggregates all derived GAP values into a single output file (GAP.csv), which serves as a prioritized roadmap for data expansion by SMEs. In this way, the dataset-estimation methodology provides a principled mechanism for continual LLM quality enhancement, ensuring that the model progressively converges toward high-fidelity, feature-complete Oracle-to-PostgreSQL conversion.

Determining which training samples are required for the next fine-tuning iteration, the dataset-estimation pipeline (Fig. 13) integrates information from four sources:

- dataset;
- feature taxonomy;
- error analysis;
- model quality metrics.

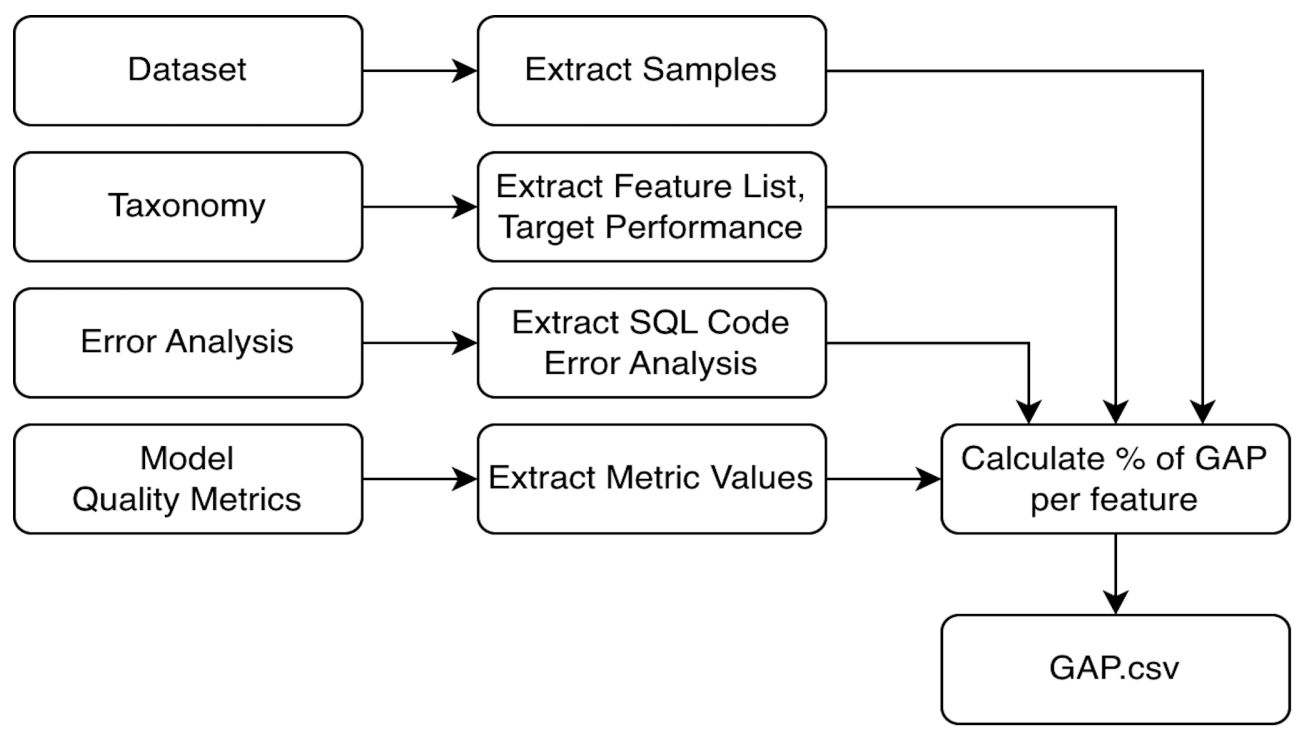

Fig. 13. Dataset estimation pipeline.

First, samples are extracted from the existing dataset, while the taxonomy provides the list of SQL features and their target performance thresholds. Error analysis contributes detailed information on SQL-level mistakes and quality metrics supply per-feature scores (Recall, BLEU, ChrF, SER). These inputs are merged to compute feature-specific performance gaps, identifying where the model underperforms.

The system then extracts the exact samples needed to close these gaps and outputs them in the GAP.csv file. This file guides the next round of data collection or SME annotation enabling automated and targeted improvement of LLM migration quality across iterations.

## III. Results and Discussion

To systematically evaluate the performance of LLMs during migration from Oracle to PostgreSQL, we considered many different candidate systems, including Qwen32B-base, Qwen32B-ft2, Qwen32B-ft1, Qwen7B-ft2, GPT-4.1-mini, GPT-4.1-nano, GPT-4o-mini, GPT-4o and traditional migration tool Ora2PG. To account for the heterogeneity of real-world SQL codebases, the evaluation set was stratified by file size. Scripts were categorized into:

- large (L-size, >200 lines; 165 files);
- medium (M-size, 101–200 lines; 212 files);
- small (S-size, ≤100 lines; 1,425 files).

This stratification enables analysis of model behavior under different structural and semantic complexities.

### A. Conversion pipeline

The chart (Fig. 14) summarizes the translation quality of all candidate models based on NLP (Recall, BLEU and ChrF) in the translation process. The Qwen32B Tuned model demonstrates a significant advantage across all metrics, consistently outperforming all other language models and the traditional Ora2PG tool. Ora2PG performs the worst across all categories, confirming that language model-based significantly outperforms direct rule-based script-to-script conversion.

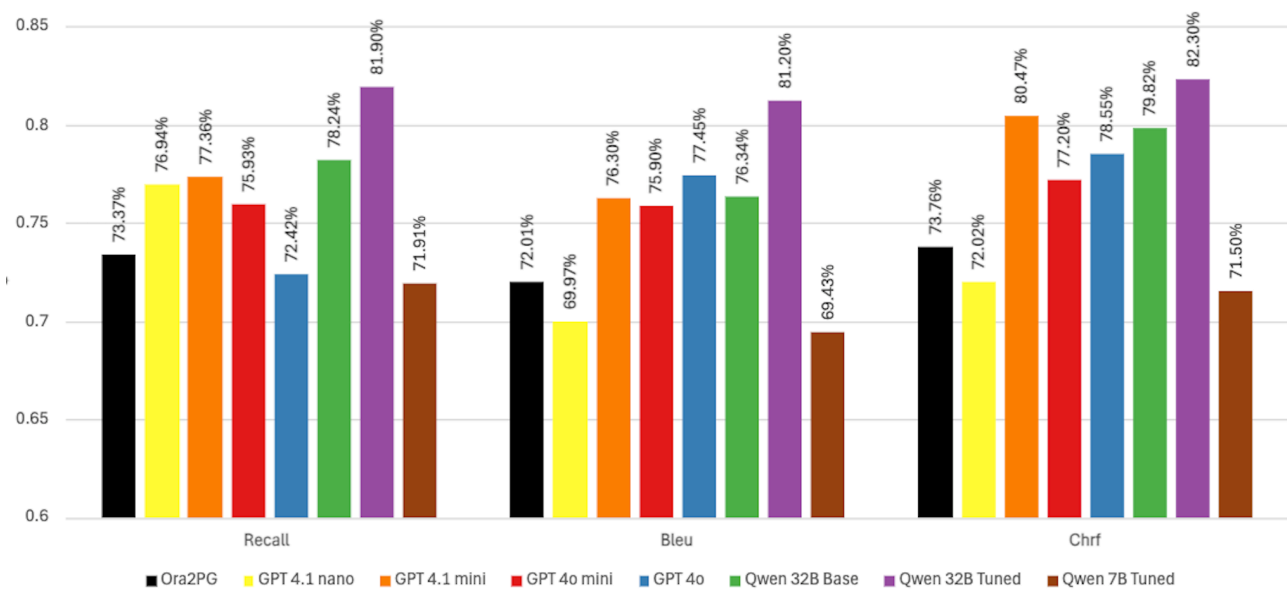

Fig. 14. Quality of conversion pipeline.

Across most models, BLEU values cluster around 0.6-0.8, indicating generally high-quality conversions even without retrieval augmentation or additional context. Both Recall and ChrF exhibit similar trends, further validating the reliability of LLM-generated PostgreSQL output. This demonstrates the effectiveness of two-stage fine-tuning and feature-aware dataset refinement in enhancing conversion precision and consistency.

## B. History-aware conversion pipeline

This chart (Fig. 15) presents NLP metrics for the history-aware conversion pipeline, where previously converted chunks are included as contextual input. In this configuration, Qwen32B Base demonstrates the strongest and most stable overall performance across Recall, BLEU and ChrF, outperforming all other LLM candidates. As in the conversion pipeline, Ora2PG is consistently inferior to every LLM-based variant.

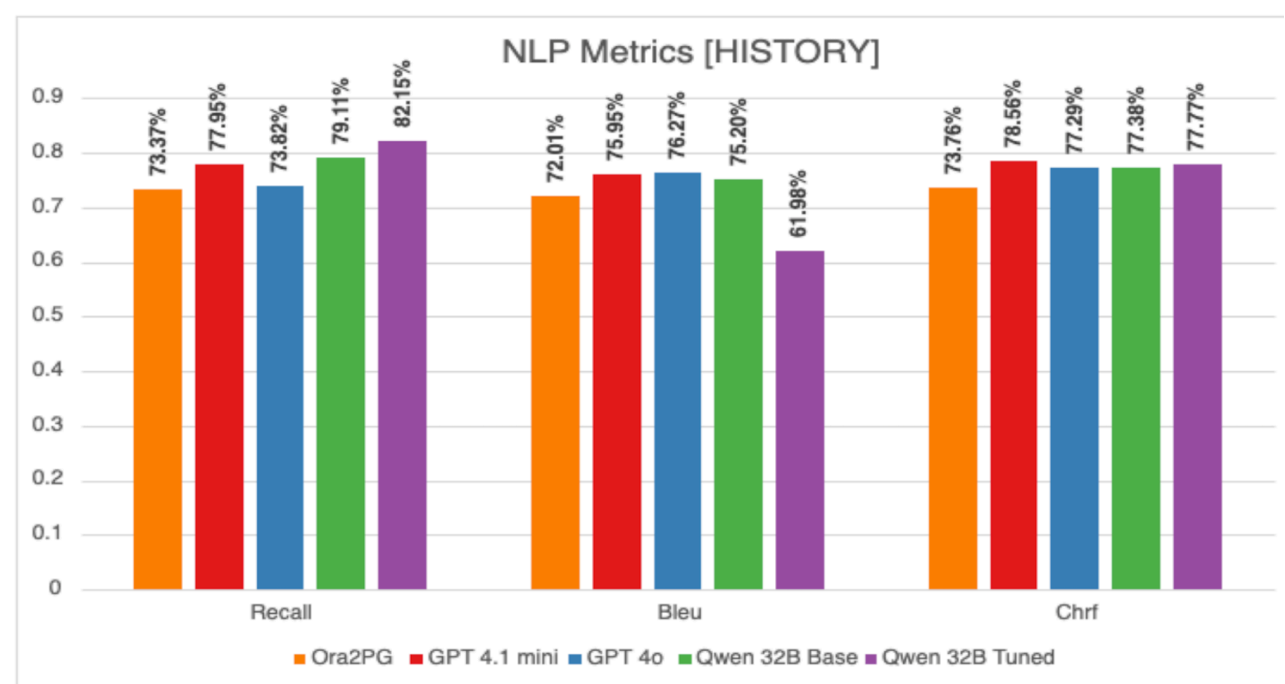

Fig. 15. Quality of history-aware conversion pipeline.

Across all models, the BLEU score remained around 0.6 or higher, indicating that the overall conversion quality remained high even with the introduction of historical information. However, the Qwen32B Tuned model had the lowest BLEU score among all models, suggesting that the fine-tuned model was more sensitive to additional historical context and required specific adjustments to reduce perplexity.

The GPT and Qwen32B Base models exhibit highly similar metric characteristics, demonstrating their robustness in handling history-introduced sequence dependencies. These results collectively highlight the excellent performance of the base LLM model in history-aware environments and underscore the necessity for targeted fine-tuning of models that need to operate within extended historical contexts.

## C. Conversion pipeline with RAG

The chart (Fig. 16) presents NLP metrics for the RAG-based conversion pipeline under different strategy configurations.

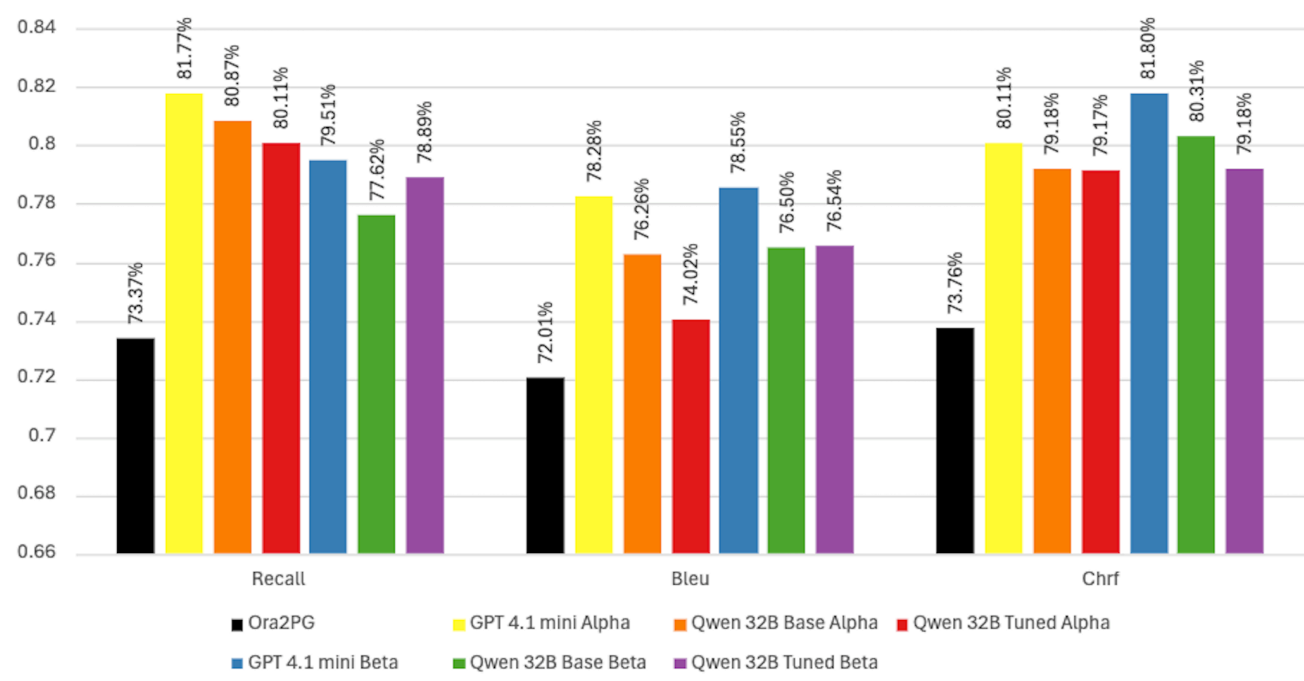

Fig. 16. Quality of history-aware conversion pipeline.

Overall, all LLM candidates exhibit comparable performance, with only marginal differences across Recall, BLEU and ChrF. The Alpha configurations, both for GPT-4.1-mini and Qwen32B, show a slight but consistent advantage over their corresponding Beta variants, indicating that certain retrieval or prompt-augmentation settings yield more stable contextual grounding.

As with previous approaches, Ora2PG significantly underperformed all LLM-based systems, confirming its limitations in retrieving enhanced neural transfer. Overall, the results demonstrate that RAG exhibits superior performance across all models with minimal sensitivity to model series or policy versions.

## D. Metrics aggregation

Aggregation of rest experimental results, including syntax correctness is shown in Table 3. The table reports the performance gain of each model-pipeline configuration relative to Ora2PG, which serves as the baseline migration tool. Two types of aggregate scores were computed: Full Set of Metrics (FSM) and Custom Metrics (CM).

FSM reflects improvements across the entire set of evaluation metrics (Recall, BLEU, ChrF, normalized syntax error rate, normalized warnings, and other diagnostic indicators).

CM, NLP-based metrics (Recall, BLEU and ChrF), represents improvements computed using a custom, reduced subset of metrics, allowing sensitivity analysis for scenarios where only selected quality dimensions are prioritized. In addition, we assigned extra weight to each metric derived from business objectives - trade-off between manual coding and code generation approach.

For each configuration, we first aggregated the metrics into a single composite score and then calculated the difference between the LLM result and the corresponding Ora2PG score. Positive values indicate that the LLM outperformed Ora2PG, while negative values reflect underperformance.

The results demonstrate that nearly all LLM-based pipelines substantially exceed Ora2PG, with the largest gains observed for Qwen32B Tuned in the Conversion and RAG pipelines. Even lighter models such as GPT-4.1-mini and GPT-4o-mini show significant advantages. Only isolated cases, such as Qwen7B Tuned in Conversion, show minor negative gains.

TABLE III. AGGREGATED RESULTS OF EXPERIMENTS

| N | Model, Pipeline, Strategy A/B | FSM, % | CM, % |
| --- | --- | --- | --- |
| 1 | CONVERSION Qwen 32B Tuned | 16.93 | 8.75 |
| 2 | RAG GPT 4.1 mini Alpha | 15.52 | 7.35 |
| 3 | RAG GPT 4.1 mini Beta | 15.22 | 6.59 |
| 4 | RAG Qwen 32B Base Alpha | 14.09 | 6.07 |
| 5 | RAG Qwen 32B Tuned Beta | 13.76 | 5.17 |
| 6 | RAG Qwen 32B Tuned Alpha | 13.45 | 4.94 |
| 7 | RAG Qwen 32B Base Beta | 13.28 | 4.72 |
| 8 | CONVERSION Qwen 32B Base | 13.23 | 4.89 |
| 9 | CONVERSION GPT 4o | 11.98 | 2.14 |
| 10 | CONVERSION GPT 4.1 mini | 11.77 | 4.55 |
| 11 | CONVERSION GPT 4o mini | 11.58 | 3.15 |
| 12 | HISTORY Qwen 32B Base | 11.35 | 4.54 |
| 13 | HISTORY GPT 4.1 mini | 11.31 | 4.40 |
| 14 | HISTORY Qwen 32B Tuned | 11.08 | 1.71 |
| 15 | CONVERSION GPT 4.1 nano | 9.72 | 0.81 |
| 16 | CONVERSION Qwen 7B Tuned | 7.51 | -1.96 |
| 17 | HISTORY GPT 4o | 5.25 | 2.23 |
|  | … |  |  |

Overall, the gain analysis confirms that LLM-driven migration consistently provides higher translation quality than the traditional rule-based tool.

## E. *Error groups rationale*

Quantitative assessments show that the number of errors in all pipelines has been significantly reduced. However, for better interpretation, error types are explicitly categorized and associated with specific architectural components of the proposed framework.

Syntax errors include malformed SQL queries, incorrect PL/pgSQL code block separators, missing keywords, or invalid function signatures. These errors are primarily caused by differences in Oracle and PostgreSQL syntax rules. By improving word segmentation and dialect-aware grammar patterns, such errors can be significantly reduced in Oracle-to-PostgreSQL topic-specific corpora.

Semantic errors occur when the generated code has valid syntax but altered behavior, such as incorrect NULL handling, misunderstanding of transaction range, or unusual logical deviations. Although these errors occur infrequently, they are more critical because they affect the correct execution of the program. These issues can be partially addressed through fine-tuning and the introduction of comprehensive quality metrics. Furthermore, they remain key targets for future execution-oriented validation efforts.

A missing function error occurs when Oracle-specific structures (such as higher-order PL/SQL functions, RMAN operations, or SQL*Plus commands) are omitted or oversimplified in the output. These errors are closely related to insufficient contextual information. Search Improved Generation (RAG) elements significantly improve coverage by directly addressing the issue of missing features by introducing feature-specific documents and historical paradigms into the generation context.

Structural errors refer to problems such as broken file dependencies, incorrect object creation order, or incomplete logical migrations of multiple files. These types of errors typically occur in large systems, rather than isolated scripts. History-based pipelines significantly reduce structural errors by preserving document-to-document and temporal relationships between database objects.

## F. *Results of error analysis*

We carefully analyzed the errors that were made during the conversion. These errors could include incorrect code, loss of context, etc. The four charts (Fig. 17) present the average file-conversion efficiency for each model under the Conversion, history-aware conversion pipeline (History), RAG with Strategy A (Retrieval Alpha) and RAG with Strategy B (Retrieval Beta) pipelines. Efficiency reflects the proportion of files successfully converted without critical syntax errors or structural degradation.

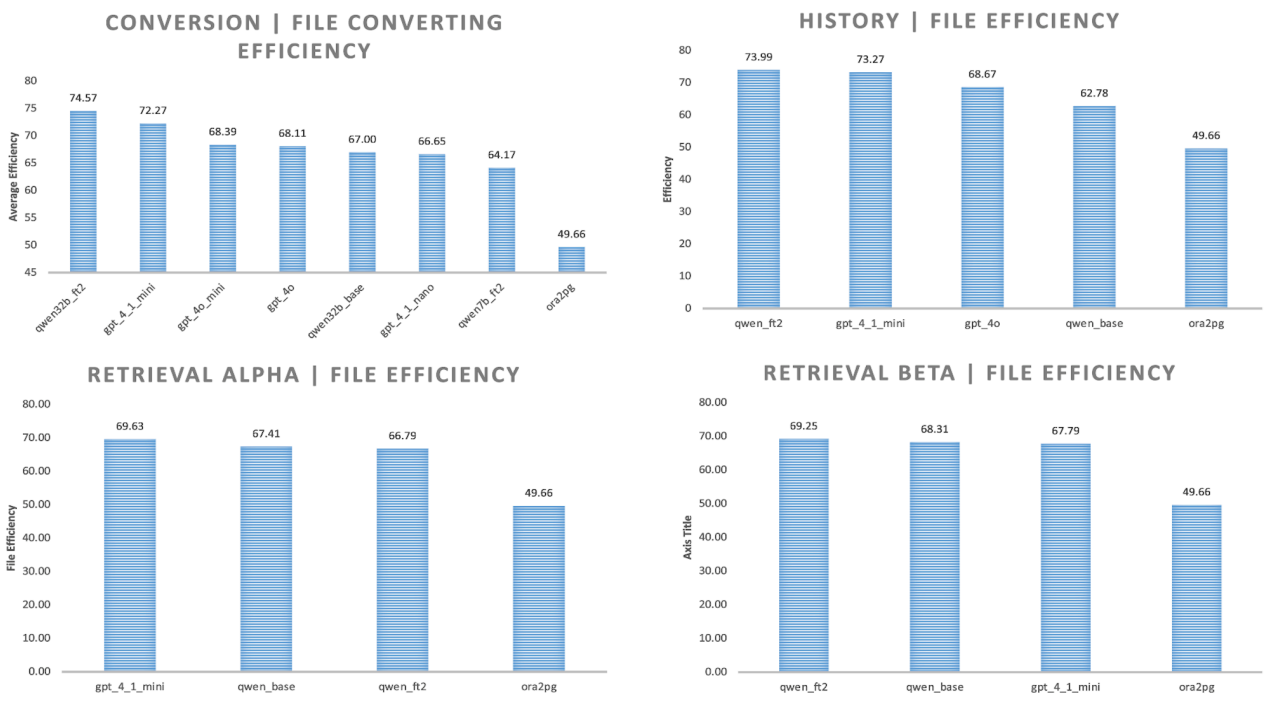

Fig. 17. File efficiency conversion by 4 pipelines.

Conversion pipeline showed that Qwen32B-ft2 achieves the highest efficiency (74.57%), outperforming all other LLMs and vastly surpassing Ora2PG (49.66%). GPT-4.1-mini and GPT-4o-mini also perform strongly. This confirms that fine-tuned LLMs deliver superior direct translation reliability.

History-aware pipeline introducing previously generated context improves consistency. Qwen32B-ft2 (73.99%) and GPT-4.1-mini (73.27%) perform nearly identically and significantly better than Qwen-base (62.78%) and Ora2PG. History provides gains for most models, especially medium and large files.

RAG with Strategy A pipeline improves structural grounding. GPT-4.1-mini leads (69.63%), followed by Qwen-base (67.41%) and Qwen-ft2 (66.79%). All LLMs substantially outperform Ora2PG. Retrieval helps but introduces slight variability due to retrieved context.

RAG with Strategy B pipeline efficiency is more balanced. Qwen-ft2 (69.25%), Qwen-base (68.31%), and GPT-4.1-mini (67.79%) show similar performance. This configuration provides strong results across models with minor differences.

Across all pipelines, Ora2PG consistently ranks last, confirming the superiority of LLM-based translation. The highest stability is observed with Qwen32B-ft2, while RAG pipelines offer additional but model-dependent benefits. By error files, Qwen32B-ft2 produces the fewest error-containing files (205 files), showing the highest stability (Fig. 18). GPT-4o (275) and GPT-4.1-nano (282) follow, while Qwen7B-ft2 (353) and GPT-4o-mini (435) perform worse. GPT-4.1-mini exhibits the largest number of failing files (481), confirming its instability in this pipeline. However, the pattern intensifies when counting total errors. Qwen32B-ft2 (3554) again leads in quality, while GPT-4.1-mini and GPT-4o-mini accumulate dramatically more errors (10008 and 8706, respectively). Qwen32B-base yields the highest error accumulation (10500), reflecting sensitivity to large procedural logic without fine-tuning.

Ora2PG, although showing fewer total errors (302), does so only because it produces short, incomplete translations, not because of correctness. Its low coverage distorts this metric. Also, we will show later, Ora2PG, failed a lot of files. That's why now it has a good value.

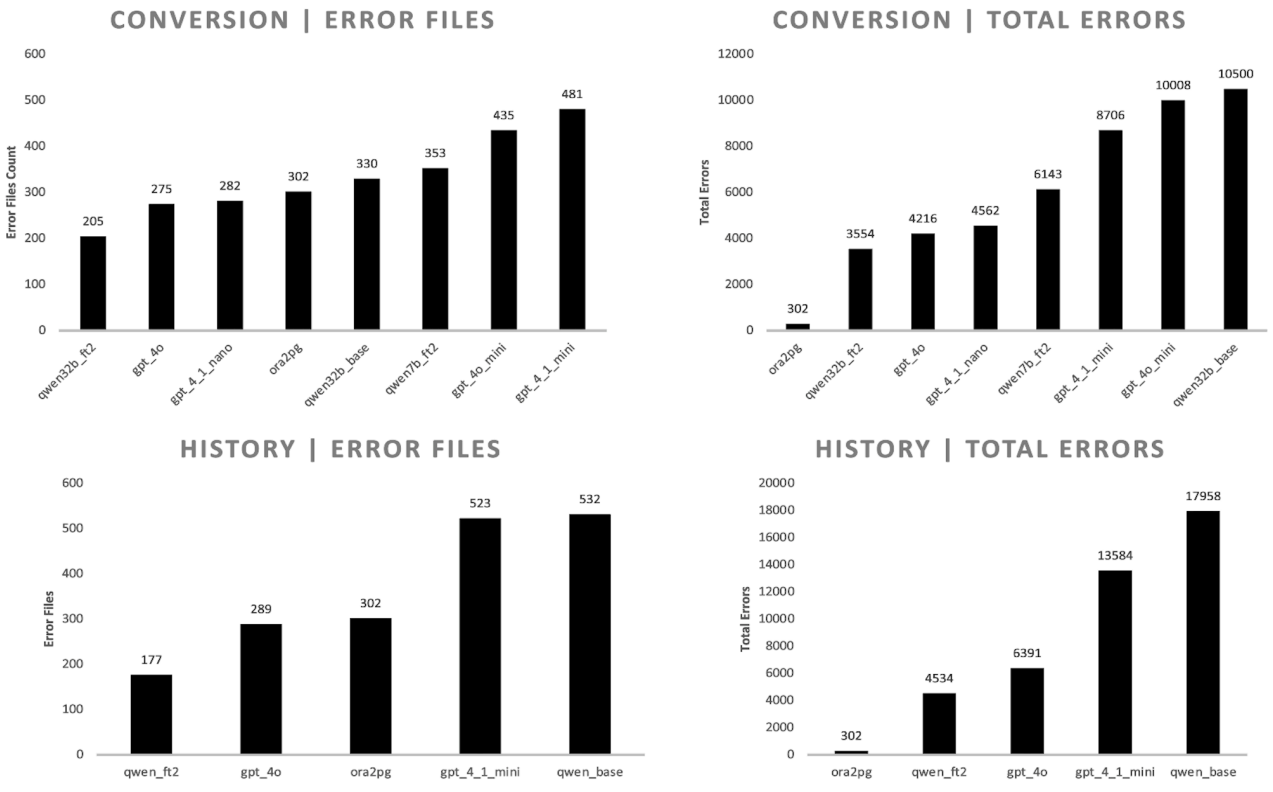

Fig. 18. Error files conversion by Conversion and History pipelines.

History-aware conversion significantly reduces errors for strong models. Qwen32B-ft2 (177) outperforms all others, followed by GPT-4o (289). GPT-4.1-mini (523) and Qwen32B-base (532) show high error incidence, indicating that added history amplifies inconsistencies for weaker or non–fine-tuned models. The same pattern holds for total errors. Qwen32B-ft2 (4534) has the lowest count among LLMs. GPT-4o (6391) is moderately stable, whereas GPT-4.1-mini (13584) and Qwen32B-base (17958) accumulate extremely high error totals, demonstrating compounding error propagation under historical context. Again, Ora2PG appears low (302) only because its incomplete outputs mask true error volume.

The same analysis we conducted for other pipelines. We analysed: syntax error rate, syntax error per line, percent of not converted scripts, SQL feature coverage, valid files by size, and coverage by size. Fig. 19 summarizes the best-performing model-pipeline combinations relative to Ora2PG and quantifies the gains achieved through fine-tuning.

Top candidates for each pipeline in comparison to ora2pg

| Model | Pipeline | File Efficiency | Class Efficiency | Size Efficiency | SER DB | Error Files | Total Errors | Not Converted Files |
|---|---|---|---|---|---|---|---|---|
| qwen32b_ft2 | Conversion | 74.57 | 63.01 | 55.87 | 11.38 | 205 | 3554.00 | 28 |
| qwen32b_ft2 | History | 73.99 | 62.85 | 53.45 | 9.82 | 177 | 4534.00 | 22 |
| gpt_4_1_mini | Retrieval Alpha | 69.63 | 60.03 | 51.90 | 12.60 | 227 | 9974.00 | 102 |
| qwen32b_ft2 | Retrieval Beta | 69.25 | 58.46 | 50.85 | 13.71 | 247 | 4157.00 | 62 |
| ora2pg | None | 49.66 | 48.21 | 42.15 | 16.76 | 302 | 302.00 | 504 |

Fine tuned vs Base improvement

| Pipeline | File Efficiency ↑ | Class Efficiency ↑ | Size Efficiency ↑ | SER_DB ↓ | Error Files ↓ | Total_Error s ↓ | SEPL ↓ | Not Converted Files ↓ |
|---|---|---|---|---|---|---|---|---|
| History | 11.21 | 12.89 | 11.85 | -19.70 | -355.00 | -13424.00 | -0.10 | 11.00 |
| Retrieval Alpha | -0.62 | 0.35 | 1.31 | -3.05 | -55.00 | -3817.00 | -0.03 | 46.00 |
| Retrieval Beta | 0.94 | -0.21 | -0.49 | -2.89 | -52.00 | -3004.00 | -0.02 | 18.00 |
| Conversion | 7.57 | 6.46 | 7.02 | -6.94 | -125.00 | -6946.00 | -0.05 | -14.00 |

Fig. 19. Summary of error analysis.

The upper table highlights that, for the evaluated test set, the Conversion and History pipelines consistently deliver the strongest results across all metrics, including file efficiency, class efficiency, size efficiency, syntax-error rate (SER_DB), and total error count. In both pipelines, Qwen32B-ft2 is the top performer, achieving the lowest number of error files and minimal non-converted files.

The lower table illustrates the direct improvements gained by fine-tuning over base models. Fine-tuning yields substantial reductions in error files and total errors, most notably in the History pipeline, where total errors decrease by over 13,000 and SER_DB drops by 19.7 points. Conversion also shows clear benefits with higher file efficiency (+7.57) and lower syntax error rate.

Overall, Fig-18 demonstrates that fine-tuning greatly enhances reliability, and that Conversion and History pipelines are the most effective strategies for high-quality Oracle-to-PostgreSQL migration.

### G. *Correlation between expected and generated features*

Correlation between predicted and generated PostgreSQL feature distributions plays a crucial role for entire migration process estimation at an earlier stage as possible because real enterprise projects have 100+ thousands scripts.

In Fig. 20, the results of analysis is shown for the base Qwen32B model (without fine-tuning). These plots clearly demonstrate that the base model exhibits less accurate feature distribution, including more pronounced under-generation of PL/pgSQL and PSQL constructs and higher variance in complex feature categories. This confirms that fine-tuning significantly improves the alignment of generated PostgreSQL code with the expected feature profile.

Fig. 21 evaluates the degree of agreement between the PostgreSQL code generated by the Qwen32B-ft2 model and the expected feature distribution derived from the general PostgreSQL feature model.

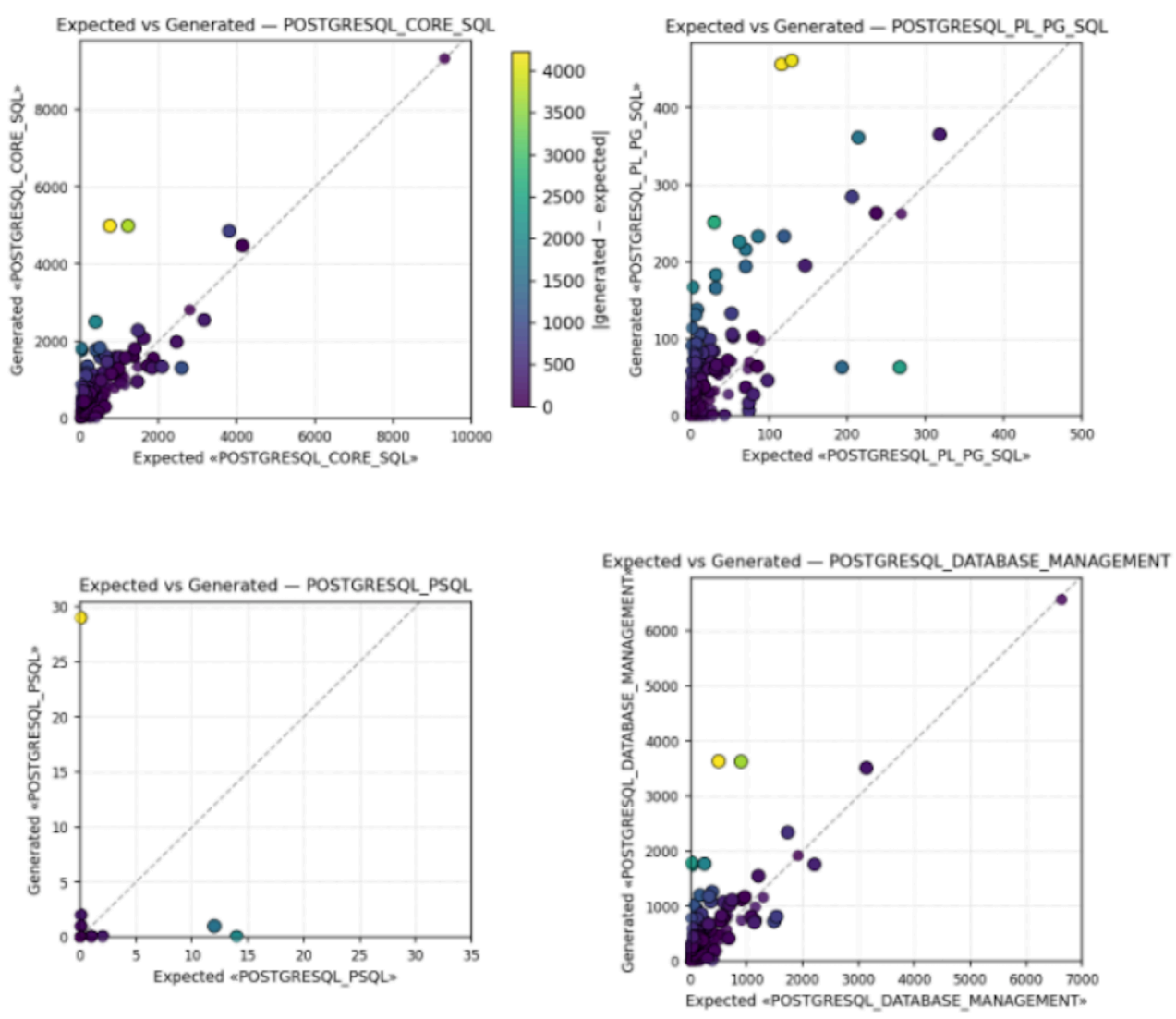

Fig. 20. Correlation between expected and generated features for Qwen32b base.

Each subplot compares expected vs. generated counts for a specific feature class: CORE_SQL, PL/pgSQL, PSQL, and DATABASE_MANAGEMENT. Points lying near the diagonal indicate strong agreement between predicted and produced feature frequencies.

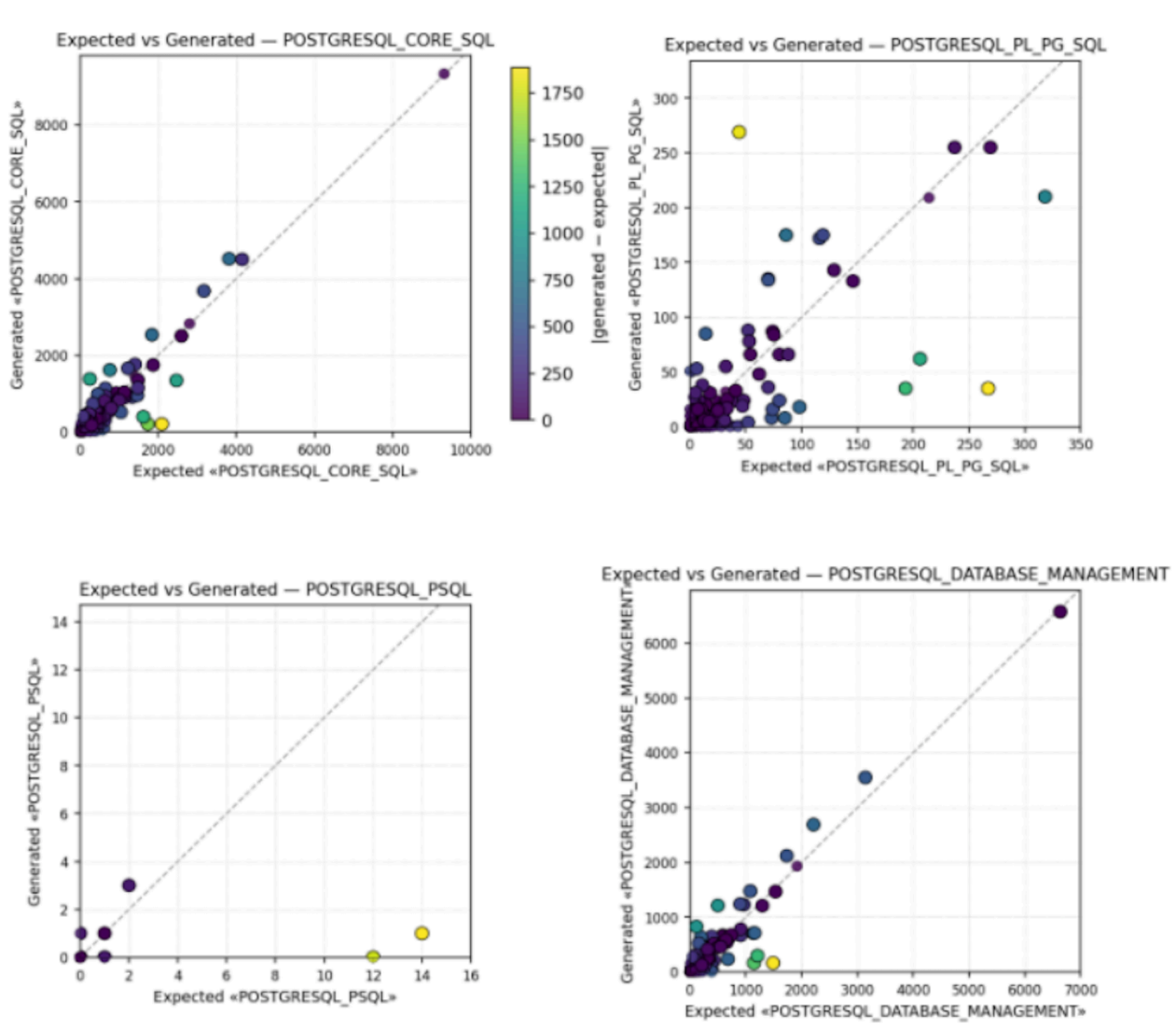

Fig. 21. Correlation between expected and generated features for Qwen32b-ft2.

The generated outputs for CORE_SQL show a strong positive correlation with expected counts. Most points cluster close to the diagonal, indicating that the model typically produces the correct volume of SQL Engine constructs. A few outliers (yellow and green points) reflect moderate overgeneration, but overall alignment is high.

For PL/pgSQL, the correlation is weaker: many samples lie below the diagonal, indicating underproduction of procedural elements relative to expected distributions. This reflects the model's conservative behavior when converting control structures, dynamic SQL or subprogram logic.

Counts of PSQL are very small, and the scatter shows high variance relative to scale, meaning that the model inconsistently generates PSQL meta-commands. Several samples significantly underproduce expected PSQL items, suggesting limited confidence or insufficient training examples in this category.

The DATABASE MANAGEMENT feature group shows the strongest and most stable correlation after CORE_SQL. Generated feature counts align closely with expectations, particularly for initialization parameters and catalog-related operations. Only minor deviations appear in high-count regions.

The expected-vs-generated scatter plots for Ora2PG (Fig. 22) reveal a substantially weaker correlation across all PostgreSQL feature groups when compared with the fine-tuned Qwen32B-ft2 (Fig. 21). While Qwen32B-ft2 generally aligns points close to the diagonal, indicating that the generated feature counts approximate the expected distribution, Ora2PG exhibits large deviations, sparse alignment, and systematic underproduction.

Qwen32B-ft2 for CORE_SQL shows a dense cluster near the diagonal, demonstrating strong proportional reproduction of SQL constructs. Ora2PG, in contrast, generates much fewer CORE_SQL features than expected; almost all points lie far below the diagonal, indicating heavy information loss and incomplete translation.

Qwen32B-ft2 for PL/pgSQL produces moderate but meaningful correlation with expected procedural features. Ora2PG almost completely fails to generate PL/pgSQL structures: points cluster near zero despite high expected values. This reflects Ora2PG's inability to convert PL/SQL logic or recreate procedural semantics.

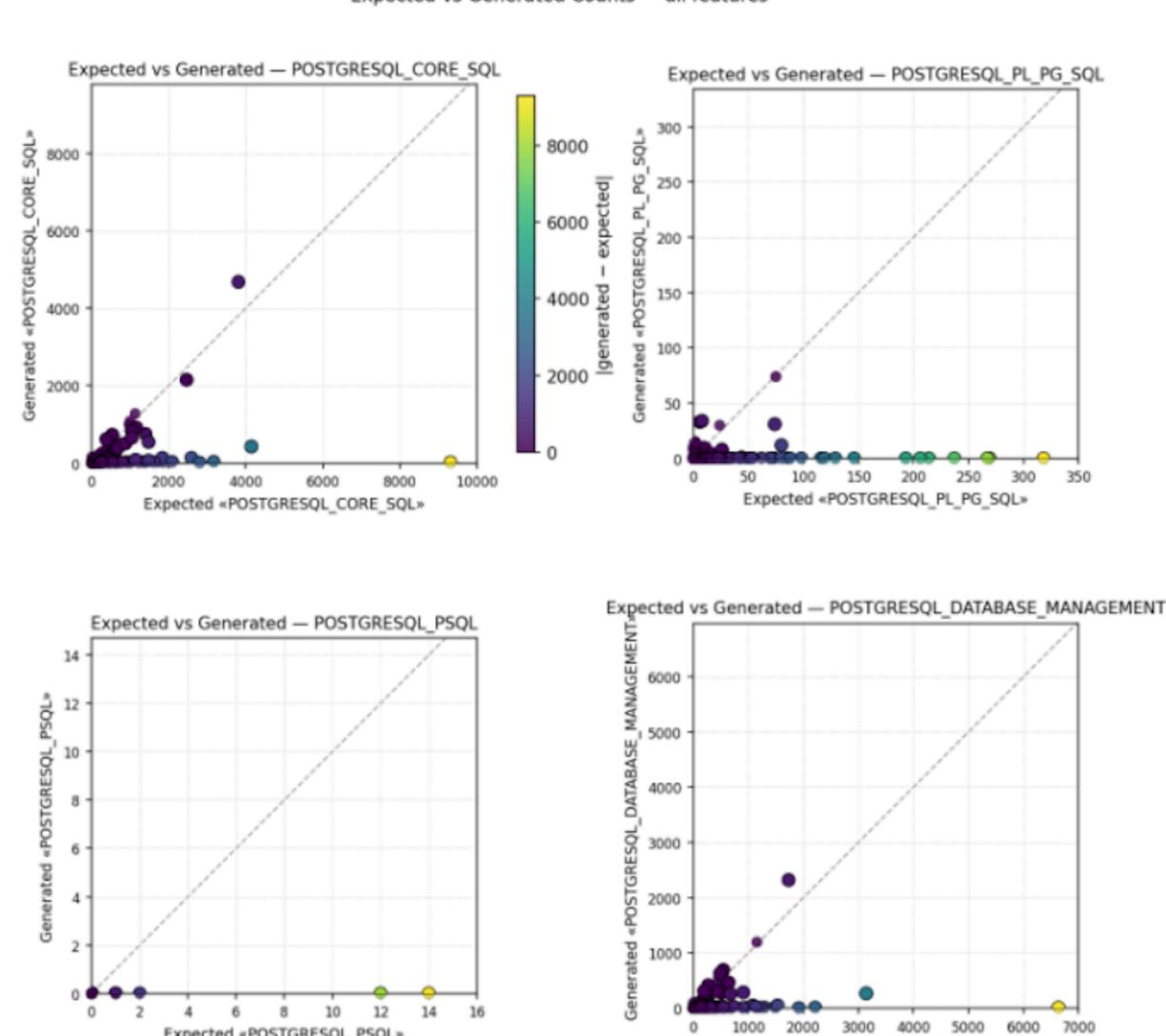

Fig. 22. Correlation between expected and generated features for Ora2PG.

PSQL for both systems show low counts due to limited dataset representation, but Qwen32B-ft2 still aligns closer to the diagonal. Ora2PG systematically undergenerates PSQL commands, indicating limited capability in handling SQL*Plus - PSQL environment commands.

DATABASE MANAGEMENT for Qwen32B-ft2 demonstrates strong correlation and preserves initialization parameters, catalogs, and monitoring constructs. Ora2PG severely underproduces this category, showing almost no alignment with expected distributions. The points cluster tightly near zero, even when expected counts are large.

This comparison highlights the fundamental limitation of rule-based migration: Ora2PG fails to preserve the structural and semantic richness of the input code, whereas the fine-tuned Qwen32B-ft2 approximates the expected distribution with high fidelity and minimal loss.

### *H. Feature coverage*

Fig. 23 illustrates the coverage of expected PostgreSQL feature groups across all evaluated models. Coverage reflects the proportion of feature instances that the generated PostgreSQL scripts reproduce compared to the expected distribution derived from the ground-truth dataset. Higher coverage indicates stronger completeness and better preservation of semantic and structural information.

CORE_SQL for most models achieve strong coverage (≈66–72%), with Qwen32B-ft2 demonstrating the highest value (72.3%). Ora2PG shows the lowest coverage (49.7%), confirming substantial information loss in rule-based translation.

DATABASE MANAGEMENT coverage is consistently high across models (≈61–65%). Qwen32B-ft2 again leads (65.4%), while Qwen7B-ft2 and Ora2PG underperform. LLMs retain most catalog, initialization, and metadata-related constructs.

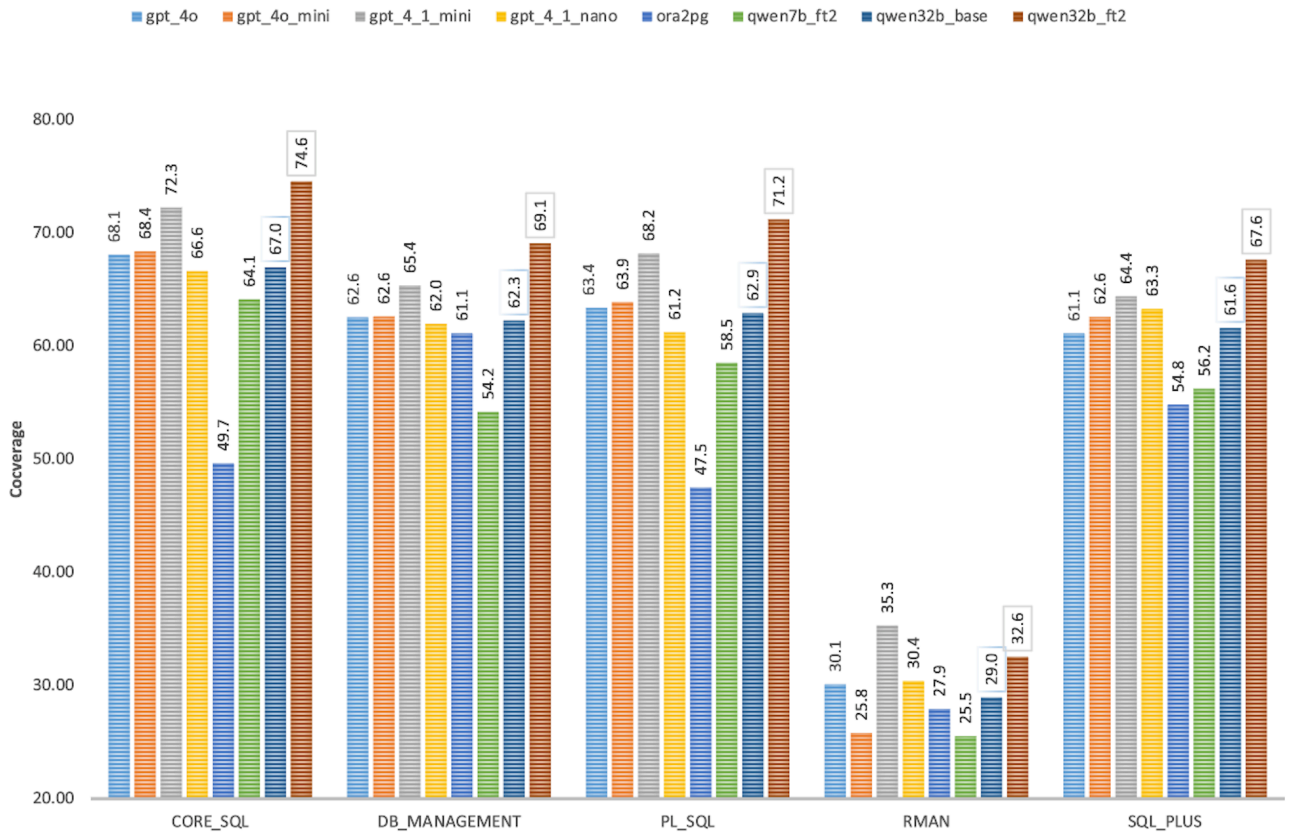

Fig. 23. Feature coverage by LLMs.

PL/SQL → PL/pgSQL coverage values are slightly lower (≈58–69%), reflecting the complexity of procedural logic conversion. Qwen32B-ft2 reaches the best coverage (69.1%), outperforming GPT models and base variants. Ora2PG shows the lowest result (47.5%).

RMAN for all models show significantly reduced coverage due to low dataset representation. Qwen32B-ft2 achieves the highest score (35.3%), while other LLMs remain near 25–30%. RMAN remains the hardest class for all systems.

Finally, SQL*Plus coverage is moderate (≈56–67%). Qwen32B-ft2 again demonstrates the strongest capability (67.6%), showing improved handling of environment commands, formatting, and session-control statements compared with other models and Ora2PG.

### *I. Rule-based tools (Ora2PG) fail at scale*

Rule-based migration tools such as Ora2PG exhibit fundamental limitations when applied to large-scale, heterogeneous Oracle codebases. The empirical results presented in the Fig. 24-26 clearly demonstrate these systemic limitations when applied at large scales and for heterogeneous SQL feature sets.

First, coverage of core SQL code (Fig. 24) shows that Ora2PG achieves competitive or even better scores only in narrowly defined syntactic subsets such as DML. However, this apparent advantage disappears when semantic breadth is taken into account. In DDL and complex DML constructs, Ora2PG exhibits high variance and under-coverage, while fine-tuned LLMs maintain stable and balanced performance across all SQL classes. This confirms that rule-based approaches are fragile outside of predefined grammar paths.

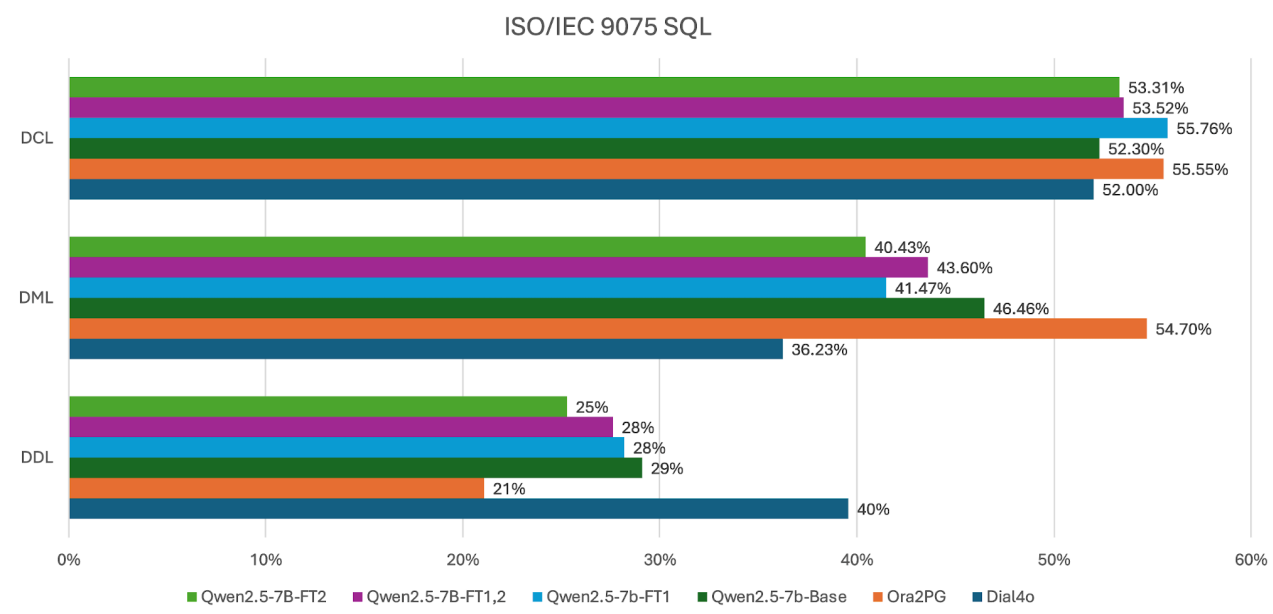

Fig. 24. SQL core coverage.

Second, the conversion pair "PL/SQL - PL/pgSQL" diagram reveals a critical semantic gap (Fig. 25). Ora2PG systematically lags behind in subroutines, such as dynamic SQL, collections and control statements. It means precisely the constructs that encode business logic.

While error handling and triggers show moderate parity, deeper procedural semantics deteriorate dramatically. In contrast, fine-tuned LLMs show consistent improvements across all PL/SQL categories, indicating better abstraction and semantic transfer rather than superficial translation.

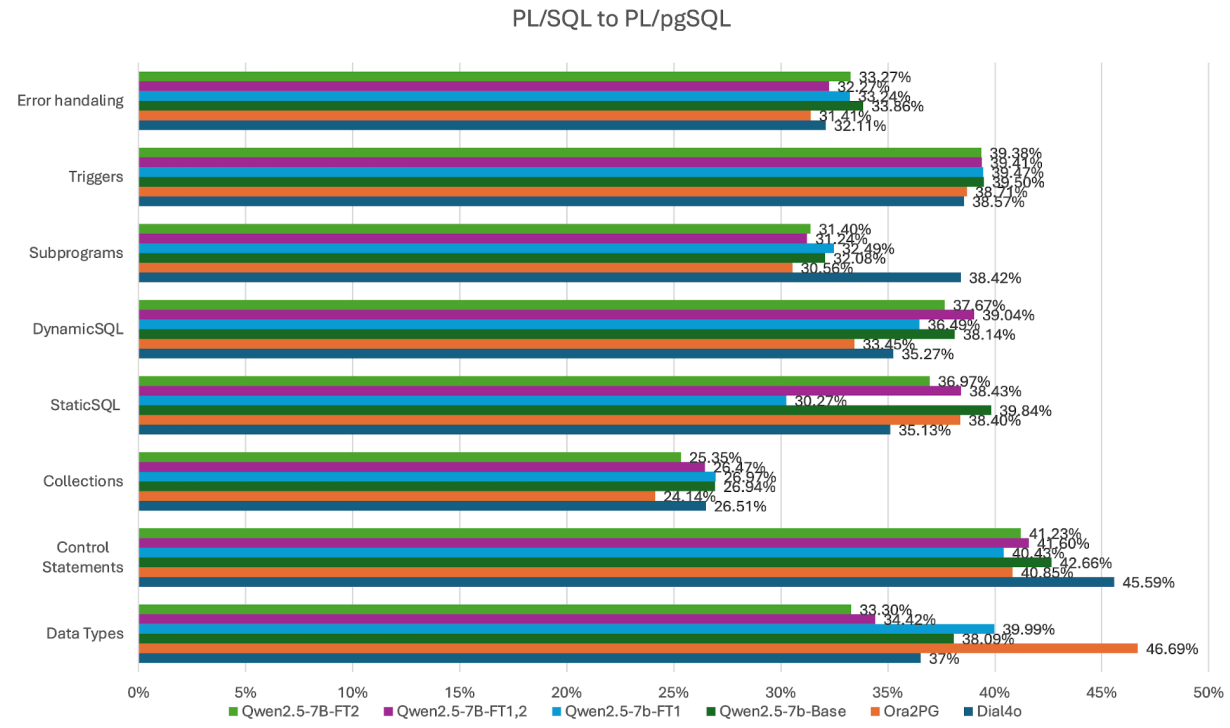

Fig. 25. PL/SQL coverage.

Third, the conversion pair "SQL*Plus - PSQL" results (Fig. 26) show almost zero effective coverage for Ora2PG in several operational categories:

- session handling;
- execution parameters;
- input/output format.

These failures do not show up as high error counts, but rather as missing or empty outputs, leading to deceptively "low" error rates. LLM-based pipelines, while imperfect, generate significantly more executable and semantically aligned code.

Overall, these diagrams confirm that Ora2PG failures are structural, not random:

- coverage drops with increasing semantic complexity;
- errors are masked by incomplete generation;
- there is no mechanism for iterative correction.

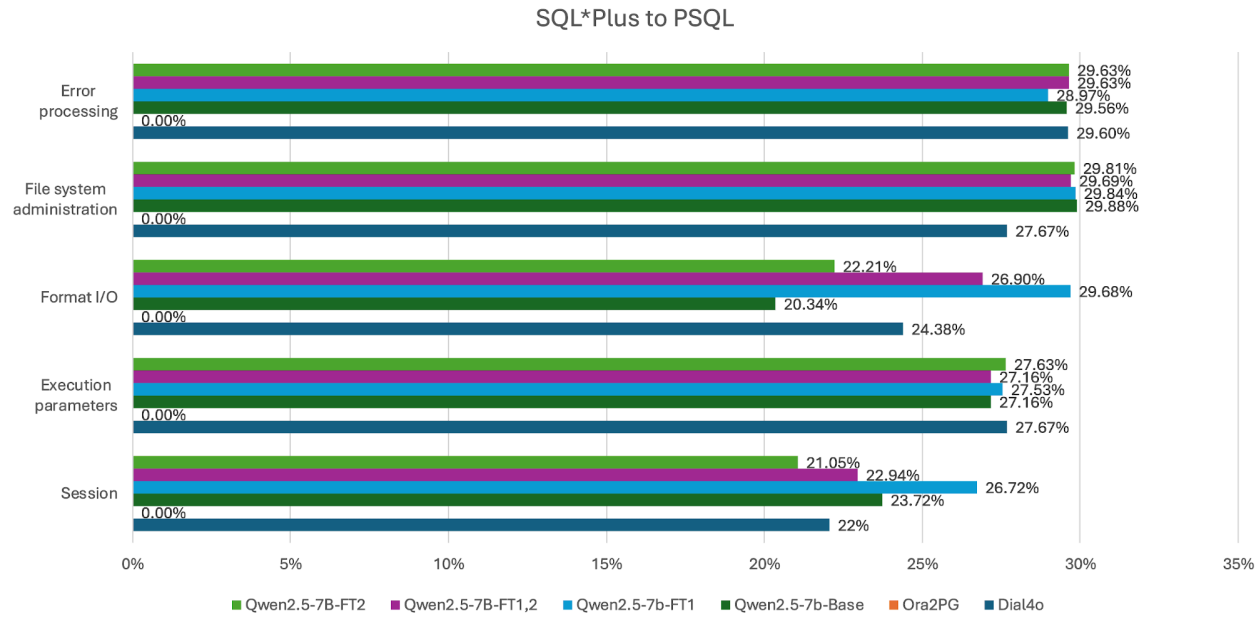

Fig. 26. SQL+ coverage.

In contrast, finely tuned LLM pipelines show gradual improvement in all features, confirming their suitability for large-scale, evolving migration scenarios.

### J. Dataset estimation for next iteration

The dataset-estimation stage is a critical component of the iterative improvement loop for model fine-tuning. Its purpose is to determine which training samples must be added, removed or rebalanced to increase the LLM's conversion quality for specific SQL features or for the overall system.

The weights and other calculated constant values were set up as it is shown in Table 4.

TABLE IV. VALUES FOR CALCULATIONS

| Weight Name | Value | Description |
|---|---|---|
| $\omega_R$ | 0,2 | Recall metric |
| $\omega_B$ | 0,2 | BLUE metric |
| $\omega_C$ | 0,2 | CHRF metric |
| $\omega_{SER}$ | 0,2 | Syntax Error Rate metric |
| $\omega_{Agg}$ | 0,4 | Aggregated metric |
| β | 0,3 | SME metric |
| $GAP_{Dict}$(CORE_SQL) | 0,0 | CORE_SQL (716920) |
| $GAP_{Dict}$(PL_SQL) | 0,53 | PL_SQL (336728) |
| $GAP_{Dict}$(SQL_PLUS) | 0,76 | SQL_PLUS (169776) |
| $GAP_{Dict}$(DM) | 0,98 | DATABASE_MANAGEMENT (13503) |
| $GAP_{Dict}$(RMAN) | 0,99 | RMAN (473) |

Using formulas (2-5) and experimental results we can calculate the gap in Oracle SQL features to make right updates on the next iteration for the predefined pipelines.

For Conversion pipeline Qwen 32B Tuned has showed, rounded after 6 digits:

- Recall=0.819011;
- Blue=0.812023;
- Chrf=0.823018;
- SER=0.882908;
- AGG=0.771871.

According to (2):

$$Q_{Raw} = 0.2R + 0.2B + 0.2C + 0.2(1 - SER) +$$

$$+ 0.4AGG = 0.2 * 0.819011 +$$

$$+ 0.2 * 0.812023 + 0.2 * 0.823018 +$$

$$+ 0.2 * 0.882908 + 0.4 * 0.771871 =$$

$$= 0,1638022 + 0,1624046 + 0,1646036 +$$

$$+ 0,1765816 + 0,3087484 = 0,9761404.$$

$$Q_{Norm} = \frac{0,9761404}{1,2} = 0,81345033.$$

According to (4), the $GAP_{Quality}$:

$$GAP_{Quality} = 1 - 0,81345033 = 0,18654967 .$$

According to (5) and Table 4, we can calculate the $GAP_{Feature}$ per each feature. The results of calculations are shown in Table 5.

TABLE V. $GAP_{Feature}$ per each feature

| Feature Name | $GAP_{Feature}$, % |
|---|---|
| CORE_SQL | 10,18 |
| PL_SQL | 36,84 |
| SQL_PLUS | 44,05 |
| DATABASE_MANAGEMENT | 49,55 |
| RMAN | 49,78 |

These GAP calculations were demonstrated for one pipeline, but the framework allows us to compute GAP values for each pipeline independently. Each pipeline uses different prompting modes and different contextual mechanisms. This means that a feature may behave differently depending on the pipeline:

- some pipelines may already perform well on PL/SQL but poorly on SQL*Plus;
- others may struggle with DB management tasks but perform better on RMAN;
- a feature with a small GAP in Conversion mode may have a large GAP in RAG mode.

By calculating GAP per feature per pipeline, we can:

1. Identify which features are bottlenecks for specific usage modes. History pipeline may require more DB_MANAGEMENT samples than Conversion pipeline.

2. Prioritize data collection where it has the highest effect. Instead of adding all types of samples blindly, we add the ones that yield the maximum performance increase.

3. Automatically recommend which samples should be added to the dataset. For each pipeline, GAP tells us which feature to expand, and how urgently it needs new samples.

This makes the dataset improvement loop fully data-driven, automated, and targeted, ensuring fastest quality growth of the model with minimum annotation cost.

This mechanism enables two important capabilities:

- Improving weak or underrepresented features. If a feature such as PL/SQL, SQL*Plus or RMAN shows low aggregated performance or high syntax-error rate, the GAP% will highlight it. The system then identifies the exact number and type of samples needed to improve performance in the next fine-tuning iteration. This ensures targeted dataset growth, rather than blind expansion.
- Adding new SQL features. When introducing a new SQL feature that did not exist in the earlier dataset, its initial performance will be near zero, resulting in a very high GAP%. The estimation framework automatically signals the need to gather training samples for this new feature category. This allows the LLM to extend its capabilities seamlessly, supporting continuous SQL-dialect expansion.

The dataset-estimation stage serves as an automated, data-driven controller for incremental model improvement. By quantifying performance gaps at feature level and mapping them to concrete sampling requirements, it ensures that each new fine-tuning iteration moves the model toward higher completeness, better feature coverage and lower error rates. Ultimately, this stage enables sustained quality growth of the LLM, even as new SQL constructs or migration patterns are introduced.

Fig. 27 quantifies the projected economic benefit of adopting the LLM-based conversion pipeline instead of relying on the traditional Ora2PG tool.

| Features | Coverage, % | Files* | Description | Converting quality, 82% | High Probable Successful Conversion, files | High Probable Successful Conversion for Ora2PG, files |
| --- | --- | --- | --- | --- | --- | --- |
| **Core_SQL** | 74.6 | 28,210 | These include SQL queries and database operations that are commonly used in the application's backend. | 74.6×82=61.17 | 28,210×61.172%=17,263 | 9,345 |
| **PL_SQL** | 69.1 | 22,260 | PL/SQL (procedures, functions, triggers) often constitute a large portion of Oracle-based systems. | 69.1×82=56.66 | 22,260×56.662%=12,613 | 10,640 |
| **SQL+** | 71.2 | 36.230 | SQL+ is typically used for scripts like reports, batch processing, or database initialization steps. | 71.2×82=58.38 | 36,230×58.384%=21,157 | 12,181 |
| **DB_MAN** | 32.6 | 10,210 | Includes administrative scripts and database management operations. | 32.6×82=26.73 | 10,210×26.732%=2,729 | 3,600 |
| **RMAN** | 67.6 | 39,170 | Backup and recovery tasks typically consume a smaller portion of the total codebase. | 67.6×82=55.43 | 39,170×55.432%=21,692 | 11,998 |
| | | | | | **Total: 75,454** | **47,764** |

***Overlapping assumptions:**
Core_SQL and PL_SQL overlap: 35% of the files (26,110 files overlap). - 74,600−26,110−20,280=28,210
PL_SQL and SQL+ overlap: 30% of the files (20,730 files overlap). - 69,100−26,110−20,730=22,260
SQL+ and DB_MAN overlap: 20% of the files (14,240 files overlap). - 71,200−20,730−14,240=36,230
DB_MAN and RMAN overlap: 25% of the files (8,150 files overlap). - 32,600−14,240−8,150=10,210
Core_SQL and RMAN overlap: 30% of the files (20,280 files overlap). - 67,600−20,280−8,150=39,170

**Difference: ± 27,690 files**

**SME plays crucially important role:**
1 SQL SME provides approx. 150 samples per day

Used dataset was prepared during 6 sprints:
30K of trainset and 2K of testset

Fig. 27. Some economical calculations.

The comparison is based on five major Oracle feature groups, weighted by their estimated coverage in real industrial systems and adjusted for typical feature overlaps. Using the observed conversion quality after the first fine-tuning iteration (≈82% successful translation within each feature), the system is expected to successfully convert approximately 75,454 files. Under identical assumptions, Ora2PG is expected to convert only 47,764 files. This yields an immediate positive difference of ≈27,690 successfully converted files, representing tasks that the LLM can complete automatically but Ora2PG cannot.

This difference translates directly into substantial cost and time savings because every unconverted file must otherwise be migrated manually by a human SME. Given that a senior SQL SME produces approximately 150 high-quality migration samples per day, converting manually the missing 27.6K files would require approximately 184 person-days per SME or 37 full-time SME-weeks or 9 SME-months of manual labor.

Thus, even a single iteration of LLM fine-tuning effectively eliminates the equivalent of nearly a year of manual migration work, significantly reducing labor costs and project timelines. Furthermore, the dataset used to achieve this improvement, 30K training samples and 2K test samples, was prepared over only six development sprints by one SME, meaning that the LLM reached economically meaningful performance rapidly and with modest supervisory effort.

Recent advances in intelligent software development have pointed to several promising directions for extending the proposed migration platform. One important trend is the integration of real-time anomaly detection into the code translation process. By continuously monitoring for feature distribution deviations, dynamic syntax errors, or runtime warnings during the migration process, anomaly detection models can provide early signs of semantic drift or structural degradation before errors propagate across large codebases. This naturally aligns with the proposed symptom-based assessment and gap-based monitoring, which allows for proactive correction rather than retrospective analysis.

Another emerging direction is the application of federated and distributed learning paradigms to enterprise migration scenarios. Large organizations often have multiple disparate Oracle codebases that cannot be centrally leveraged due to security or compliance constraints. Fine-tuning the federated LLM combined with local feature analysis and aggregated GAP statistics allows the model to improve overall performance while preserving data locality. This approach allows knowledge to be transferred between organizations at the level of abstract SQL templates, rather than at the level of raw source code.

From a scalability perspective, the proposed framework is clearly designed for enterprise-level code libraries containing tens or hundreds of thousands of SQL files. By decomposing migration quality into characteristic-level metrics and combining them with pipeline-specific GAP scores, linear expansion of analysis and targeted expansion of the dataset can be achieved without exponential growth in manual workload. Unlike rule-based tools, whose performance decreases sharply with increasing semantic complexity, LLM-based methods show steady performance improvements through iterative refinement.

The return on investment (ROI) is particularly significant. Experimental results show that even a single iteration of fine-tuning eliminates the need for manual migration of approximately 27,000+ files, equivalent to several months of SME effort. As the growth of the dataset is driven by GAP-based prioritization, each additional annotation cycle yields the largest marginal improvement. This transforms migration from a labor-intensive engineering task to a data-driven optimization process, making AI-assisted migration cost-effective for long-term modernization projects.

## IV. Future Work

Despite the encouraging results, this study has several limitations that should be considered.

First, the logic validation of the migrated database at the execution level was not fully performed. Although various metrics have been used to assess syntactic correctness and semantic similarity, the actual behavior of complex PL/pgSQL structures at runtime can still reveal hidden inconsistencies, especially in transaction processing and exception propagation.

Secondly, RAG-based pipelines are highly sensitive to the quality of their embedding space and search configuration. Inaccurate embedding or suboptimal fragmentation strategies can lead to partial contextual searches, which directly impacts the quality of generating complex or inter-file dependencies.

Third, rare features such as RMAN scripts and infrequently used administrative operations are still insufficiently representative in the training data. These features require larger and more specialized datasets to achieve stable conversion quality comparable to basic SQL and PL/SQL features.

Therefore, future research will focus on expanding the proposed framework from multiple directions.

The main goal is to achieve execution-based verification, which combines static analysis with partial execution or sandbox testing to verify the behavioral equivalence between the original code and the ported code.

Another direction is to develop adaptive search depth mechanisms, where the scope and granularity of the retrieved context are dynamically adjusted based on the complexity of the features and the detected uncertainty.

The framework will also be extended through automatic classification evolution, enabling the system to detect new SQL patterns and integrate new features without human intervention.

## Conclusions

This work presented a comprehensive framework for automating Oracle-to-PostgreSQL code migration using LLMs. The proposed solution integrates dataset engineering, feature-aware static analysis, retrieval-augmented generation, fine-tuning and a robust evaluation methodology into a single iterative workflow. Three conversion pipelines were developed and analyzed: a baseline Conversion pipeline, a History-aware pipeline that maintains sequential translation context, and two RAG pipelines (Strategy A/B) that retrieve contextual information from knowledge bases with different structures. Across all experiments involving Qwen and GPT model families, LLM-based systems significantly outperformed the traditional migration tool Ora2PG.

A scalable evaluation framework was created to assess conversion quality independently from the LLM. It incorporates NLP metrics (Recall, BLEU, ChrF), syntax inspection, chunk correctness and retrieval accuracy. All results are logged in MLflow for reproducibility. Error analysis reveals that fine-tuned models, especially Qwen32B-ft2, achieve the lowest syntax error rates and highest file-conversion efficiency. Conversion and History pipelines show the strongest results across heterogeneous SQL workloads.

Feature distribution analysis, enabled by HCFPE, quantifies coverage of Oracle and PostgreSQL features. Correlation analysis demonstrates that fine-tuned models align closely with expected PostgreSQL feature distributions, unlike the rule-based Ora2PG tool, which systematically underproduces key feature groups.

To support continual improvement, a mathematically formalized GAP-based dataset estimation mechanism identifies underrepresented or low-performance features and determines the required number of new samples for the next fine-tuning iteration. This enables targeted dataset growth and efficient SME involvement, including seamless integration of new SQL features.

Finally, an economic evaluation showed that even a single fine-tuning iteration yields substantial operational benefits, approximately 27,000+ of 100,000 files successfully migrated compared to Ora2PG, equivalent to several months of manual SME effort.

Overall, this framework establishes a scientifical, practical, scalable and economically advantageous approach for enterprise-scale database modernization using LLMs.

## References

[1] PostgreSQL Global Development Group. (2025). Porting from Oracle PL/SQL (PL/pgSQL Guide §41.13). PostgreSQL Documentation. Retrieved from https://www.postgresql.org/docs/current/plpgsql.html.

[2] Darold, G. (n.d.). Ora2Pg: Migrates Oracle to PostgreSQL—Documentation. Retrieved in 2025, from https://ora2pg.darold.net/.

[3] Amazon Web Services. (n.d.). Use AWS SCT to convert the Oracle schema to PostgreSQL. Retrieved in 2025, from https://aws.amazon.com/documentation-overview/.

[4] Rozière, B., Gehring, J., Gloeckle, F., Sootla, S., Gat, I., Tan, X. E., & Synnaeve, G. (2023). Code Llama: Open foundation models for code. arXiv preprint arXiv:2308.12950. Retrieved from https://arxiv.org/abs/2308.12950.

[5] Lozhkov, A., Li, R., Ben Allal, L., Cassano, F., Lamy-Poirier, J., Tazi, N., & de Vries, H. (2024). StarCoder2 and The Stack v2: The next generation. arXiv preprint arXiv:2402.19173. Retrieved from https://arxiv.org/abs/2402.19173.

[6] Pan, R., Ibrahimzada, A. R., Krishna, R., Sankar, D., Wassi, L. P., Merler, M., & Jabbarvand, R. (2024). Lost in translation: A study of bugs introduced by large language models while translating code. Proceedings of the ACM/IEEE International Conference on Software Engineering (ICSE 2024). https://doi.org/10.1145/3597503.3639226.

[7] Cummins, C., Seeker, V., Grubisic, D., Rozière, B., Gehring, J., Synnaeve, G., & Leather, H. (2024). Meta large language model compiler: Foundation models of compiler optimization. arXiv preprint arXiv:2407.02524. Retrieved from https://arxiv.org/abs/2407.02524.